\documentclass[twocolumn]{aastex63} 

\usepackage{makeidx}
\usepackage{amssymb}
\usepackage{amsmath}
\usepackage{eurosym}

\usepackage{siunitx}
\received{December 2, 2019}
\revised{March 6, 2020}
\accepted{***}
\submitjournal{ApJ Supplement Series}

\shortauthors{Krasnoselskikh et al.}

\begin{document}

\title{Localized magnetic field structures and their boundaries in the near-Sun solar wind from Parker Solar Probe measurements}

\correspondingauthor{Vladimir Krasnoselskikh}
\email{vkranos@gmail.com}

\author[0000-0002-6809-6219]{V. Krasnoselskikh}
\newcommand{\lpceeaffil}{LPC2E, CNRS and University of Orl\'eans, 3A avenue de la Recherche Scientifique, Orl\'eans, France}
\affiliation{\lpceeaffil}
\newcommand{\SSLaffil}{Space Sciences Laboratory, University of California, Berkeley, CA 94720-7450, USA}
\affiliation{\SSLaffil}

\author[0000-0002-7653-9147]{A. Larosa}\affiliation{\lpceeaffil}

\author{O. Agapitov}\affiliation{\SSLaffil}
\author[0000-0002-4401-0943]{T. Dudok de Wit}\affiliation{\lpceeaffil}

\author{M. Moncuquet}
\newcommand{\meudonaffil}{LESIA, Observatoire de Paris-Meudon, Meudon, 92195, France}
\affiliation{\meudonaffil}

\author{F. S. Mozer}\affiliation{\SSLaffil}
\newcommand{\berkleyaffil}{Physics Department, University of California, Berkeley, CA 94720-7300, USA}
\affiliation{\berkleyaffil}

\author{M. Stevens}
\newcommand{\cambridgeaffil}{Smithsonian Astrophysical Observatory, Cambridge, MA, 02138, USA}
\affiliation{\cambridgeaffil}

\author[0000-0002-1989-3596]{S. D. Bale}\affiliation{\SSLaffil}\affiliation{\berkleyaffil}
\newcommand{\imperialcollegeaffil}{The Blackett Laboratory, Imperial College London, London, SW7 2AZ, UK}
\newcommand{\queenmaryaffil}{School of Physics and Astronomy, Queen Mary University of London, London E1 4NS, UK}
\affiliation{\imperialcollegeaffil}\affiliation{\queenmaryaffil}

\author[0000-0002-0675-7907]{J. Bonnell}\affiliation{\SSLaffil}

\author[0000-0001-5315-2890]{C. Froment}\affiliation{\lpceeaffil}

\author[0000-0003-0420-3633]{K. Goetz}
\newcommand{\minnesotaaffil}{School of Physics and Astronomy, University of Minnesota, Minneapolis, MN 55455, USA}
\affiliation{\minnesotaaffil}

\author{K. Goodrich}\affiliation{\SSLaffil}

\author[0000-0002-6938-0166]{P. Harvey}\affiliation{\SSLaffil}

\author{J. Kasper}\affiliation{\cambridgeaffil}
\newcommand{\michiganaffil}{Climate and Space Sciences and Engineering, University of Michigan, Ann Arbor, MI 48109, USA} 
\affiliation{\michiganaffil}

\author[0000-0003-3112-4201]{R. MacDowall}
\newcommand{\GSFCaffil}{Solar System Exploration Division, NASA/Goddard Space Flight Center, Greenbelt, MD, 20771, USA} 
\affiliation{\GSFCaffil}

\author[0000-0003-1191-1558]{D. Malaspina}
\newcommand{\coloradoaffil}{Laboratory for Atmospheric and Space Physics, University of Colorado, Boulder, CO 80303, US} 
\affiliation{\coloradoaffil}

\author[0000-0002-1573-7457]{M. Pulupa}
\affiliation{\SSLaffil}

\author{N. Raouafi}
\newcommand{\jhuaplaffil}{Johns Hopkins University, Applied Physics Laboratory, Laurel, MD, USA} 
\affiliation{\jhuaplaffil}

\author[0000-0003-4582-7055]{C. Revillet}
\affiliation{\lpceeaffil}

\author[0000-0002-2381-3106]{M. Velli}
\newcommand{\californiaaffil}{Institute of Geophysics \& Planetary Physics, Department of Earth, Planetary \& Space Sciences, University of California, Los Angeles, CA 90095-1567, USA} 
\affiliation{\californiaaffil}

\author{J. Wygant}\affiliation{\minnesotaaffil}

\begin{abstract}
One of the discoveries made by Parker Solar Probe during its first encounters with the Sun is the ubiquitous presence of relatively small-scale structures that stand out as sudden deflections of the magnetic field. They were named “switchbacks" since some of them show up the full reversal of the radial component of the magnetic field and then return to "regular" conditions. As a result of the processing of magnetic field and plasma parameters perturbations associated with switchbacks we distinguish three types of structures having slightly different characteristics: I. Alfv\'enic structures, where the variations of the magnetic field components take place conserving the magnitude of the magnetic field constant; II. Compressional, where the field magnitude varies together with changes of the components of the field; III. Structure manifesting full reversal of the magnetic field (an extremal class of Alfv\'enic switchback structures). Processing of structures boundaries and plasma bulk velocity perturbations lead to the conclusion that they represent localized magnetic field tubes with enhanced parallel plasma velocity and ion beta (presumably due to the hotter plasma contained in the tube) moving together with the surrounding plasma. The magnetic field deflections before and after the switchbacks reveal the existence of total axial current. The electric currents are concentrated on the relatively narrow boundary layers on the surface of the tubes and determine the magnetic field perturbation inside the tube. These currents are closed on the structure surface, and typically have comparable azimuthal and the axial components. The surface of the structure may also accommodate an electromagnetic wave, that assists to particles in carrying currents. We suggest that the two types of structures we analyzed here may represent the local manifestations of the tube deformations corresponding to a saturated stage of the Firehose instability development. The macroscopic role of the observed magnetic structures is in providing the mechanism for the dissipation of the plasma energy supplied by the jets generated in the low corona.
\end{abstract}

\keywords{solar wind, magnetic structures, MHD waves} 

\section{Introduction}   

The Parker Solar Probe (PSP) mission \citep{fox15} addresses two fundamental opened questions in space physics, which are coronal plasma heating and the acceleration of solar wind plasmas. In both problems, wave-particle interactions involving MHD waves are known to play an important role \citep{Coleman1968,belcher71,Heyvaerts83}. During its first encounter with the Sun in November 2018, PSP revealed a multitude of sudden reversals of the nearly radial magnetic field, which have been called ``switchbacks'' \citep{bale19,kasper19}. These structures have also been named ``jets'' because the (mostly radial) flow velocity is larger inside them. The typical velocity increase is of the order of local Alfv\'en velocity (about 80-100~\si{km/s}). 

The conditions of the solar wind during PSP's first  encounter were comparable to those met by the Helios A (1974-1985) and Helios B (1976-1979) missions. PSP's closest distance to the Sun was 35.7 solar radii or 0.174 \si{AU} whereas  Helios A and Helios B had perihelia of respectively 0.31 and 0.29 \si{AU}. One of the major findings of Helios was the existence of a structured and highly intermittent solar wind alternating between fast and slow streams  \citep{Neubauer&Barnstorf1981, Denskat1981, Marsch_b, Marsch_c, Marsch_a}. Initial studies of its magnetic field measurements focused on the presence of different types of discontinuities that were observed both in fast and slow solar winds. \citet{Neubauer&Barnstorf1981} and \citet{Burlaga1977} classified these boundaries as tangential discontinuities (TD) when the normal component of the magnetic field is small with respect to its magnitude, and rotational discontinuities (RD) when the normal component is of the same order of magnitude as the field. An important difference between them is that  boundaries of the TD type prevent mass, momentum or energy exchange across the boundary, whereas RD's allow such  exchanges.  It was also noted that Alfv\'enic wave activity was higher in fast than in slow wind regimes \citet{Marsch2006}. 
\citet{Marsch_d} reported that at  perihelion, during periods of low solar activity, the solar  wind velocity was  typically \SIrange{300}{400}{km/s} with occasional bursts up to 600 \si{km/s} that could  last for a few hours. The correlation between their velocity and magnetic fluctuations,  which is usually  regarded as a condition for of Alfv\'enic type fluctuations, was shown to be qualitatively verified for the discontinuities that were observed by Helios. According  to \citet{Denskat1981}  the level of Alfv\'en wave activity decreased with radial distance.

Interestingly, \citet{Marsch_c} reported the existence of sudden events with ``enormous deviations of the magnetic field elevation angles'' of up to \ang{45}. These  were associated with deflections of the direction of the velocity vector by  up to \ang{10}. These events were arguably the earliest observations of the structures that have later been named switchbacks and  have since been observed by many others  \citep{Yamauch2004, Landi2005, Landi2006, Suess2007, Gosling2009, Neugebauer2013, Matteini2014, Borovsky2016, Horbury2018} before  PSP revealed their ubiquity in the slow solar wind. The switchbacks that have been observed by PSP during its first and third encounters are most likely the same structures as those reported by \citet{Marsch_c} except that we  can now observe them at  a much  earlier stage of  their evolution in a near-pristine solar wind. 

Another  noticeable difference between Helios and PSP is the high speed of the latter, thanks to  which it  was almost co-rotating with the Sun during its first three encounters. Because of that, the temporal variations  seen by PSP  are dominated by radially-moving  structures,  and not be the spacecraft  crossing spatial inhomogeneities.

During the closest approach of the first encounter there were observed several hundred structures per day with the deviation of the magnetic field larger than 30 degrees and duration larger than several seconds. Surprisingly there were very few such structures during the second encounter. Presumably it is related with the connectivity of the satellite position during first encounter with the equatorial coronal hole.
We have analyzed in detail twenty such structures and in the following discuss three typical ones with the aim of unravelling the properties and role of their boundaries. We intentionally selected short duration events that last for a few minutes and avoided  longer but visually more complex structures that often shows  the presence of substructures. Our three examples stand  out by offering a sharp transition from a stationary solar wind to a regime with different properties, and then  back to  the  initial conditions. For each of  them we pay particular attention to the characteristics of their boundaries, which are crucial for understanding mass and energy exchanges with the surrounding plasma flow.

An important characteristic of these  structures is: are they compressible or not? The first example is purely Alfv\'enic, with a total magnetic field that is almost constant in time. 
The notion Alfv\'enic is attributed to magnetic field perturbations that satisfy two conditions. First, the variations of the magnetic field occur without change of its magnitude, and, second, the variations of the magnetic field vector and velocity vector happen simultaneously and satisfy the following relation:
\begin{equation*}
\delta \mathbf{V}= \mathbf{V}_{A} 
\frac{\delta \mathbf{B}}{B}  
\left(1-\frac{4\pi \left( p_{\parallel}-p_{\perp} \right) }{B^{2}} \right)^{1/2}.
\end{equation*}
Here $\delta V$ is the variation of the velocity vector, $V_{A}$ is the Alfv\'en speed, $ B $ magnetic field, $\delta B$ magnetic field variation, $ p_{\parallel}$ and $ p_{\perp}$ thermal ion pressure parallel and perpendicular to the magnetic field.

The second structure is compressible while the third one shows a complete reversal of the magnetic field. 

The article is organized as follows: after presenting the data in Sec.~\ref{sec:data} we present the three examples in Secs.~\ref{sec:event1}-\ref{sec:event3} and subsequently discuss them in Sec.~\ref{sec:interpretation}. Section~\ref{sec:conclusion} concludes the study.

\section{Data}
\label{sec:data}

In our study we focus on \textit{in situ} measurements of the magnetic field from the MAG fluxgate Magnetometer and the SCM
Search-Coil Magnetometer. Both are part of the  FIELDS suite \citep{Bale2016}, which is  devoted to the measurements  of electric and magnetic fields.  MAG and SCM measure three components of the magnetic field with a sampling rate of 293 samples per second. MAG, however, measures the DC field whereas the SCM measures the AC field only; the crossover between the sensitivities of the two instruments occurs between 4 and 10 \si{Hz}.

The electron density is determined from the Quasi-Thermal Noise technique (QTN) \citep{moncuquet20}, which uses the location of the plasma line in electric field spectra to infer the electron density. This technique offers the advantage of providing a  density estimate that is independent of calibrations and spacecraft perturbations. In the following we average typically few tens of spectra to reduce the noise level, so that the final time resolution of the density is 10 to 20 \si{s}. 

The proton density and velocity are derived from moments of the proton velocity distribution as provided by the Faraday cups of the  SWEAP  suite \citep{Kasper2016}. A complete scan of the velocity distribution takes 0.8 \si{s}, which sets the cadence of these measurements. Their main asset with respect to the electron density is their better time resolution; therefore we shall rely on proton data to study fast transients. In counterpart, these measurements are sensitive to the floating potential, and therefore require inter-calibrations with other instruments. The velocity distribution also gives access to what we shall loosely call an ``indication of the temperature'', which is the average variance of the thermal velocity of the ions $k_{B} T_{i}=M_{i}\left(  \delta V \right)^{2}/2$, here $k_{B}$ is the Boltzmann constant.

Throughout our  study, we express our data in RTN coordinates. $R$ points from the Sun center to the spacecraft; $T$ lies in the spacecraft plane (close to the ecliptic) and is defined as the cross product of the solar rotation axis with $R$; points in the direction of prograde rotation. $N$ completes a right-handed system.

\section{Main properties of switchbacks}
\label{sec:properties}

\begin{figure} 
	\centering
    \includegraphics[width=\linewidth]{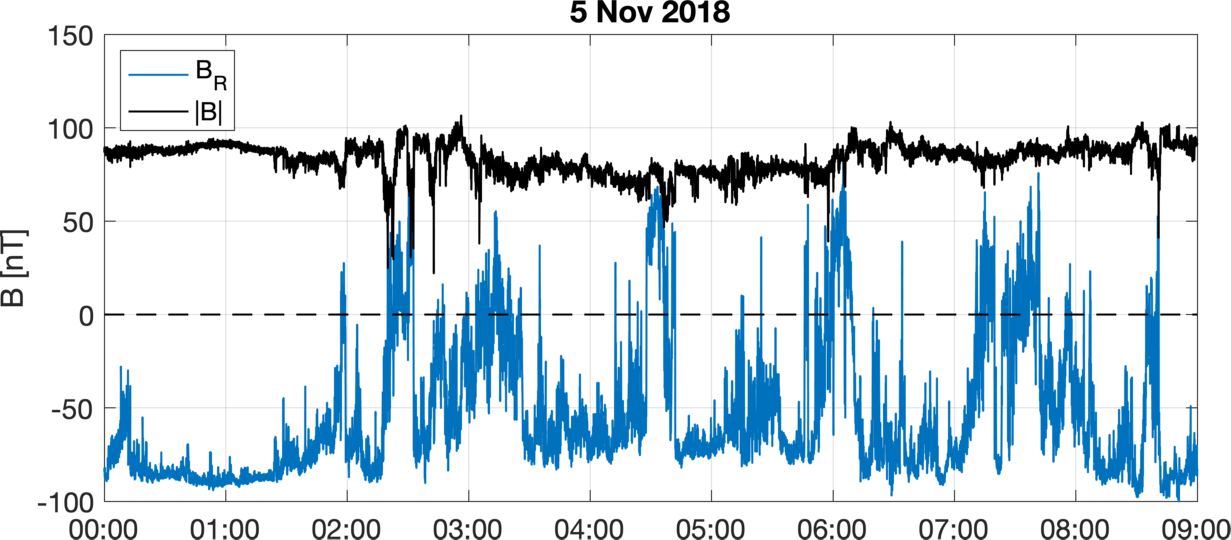} 
    \caption{Magnitude (in black) and the radial component (in blue) of the magnetic field recorded on November 5, 2018 from 00:00 to 09:00 UT.  Switchbacks show  up as sudden increases in the radial component.}
    \label{fig:fig_1}
\end{figure}

Figure~\ref{fig:fig_1} illustrates the main signature of switchbacks, which is sudden deflection of the magnetic field away from the Parker spiral. During its first encounter with the Sun, PSP observed for ten consecutive days a slow but highly Alfv\'enic solar wind stream that was originating from a small equatorial coronal hole \citep{bale19}.
The high Alfv\'enicity of the plasma is attested by the nearly constant total magnetic field, regardless of the variations of the individual components. The radial component of the magnetic field is on average negative because of the negative polarity of the field in the coronal hole. Switchbacks stand out by the rapid increase of the radial component, occasionally even leading to complete inversions of the magnetic field \citep{ddw20}. Their duration ranges from seconds to more than one hour.

In the following we shall focus on three examples that correspond to three types of structures: first, an event that manifests pure
Alfv\'enic properties; second, one that is compressional because the total magnetic field changes inside the structure; finally, a  switchback with a full inversion of the magnetic field. Following this we shall discuss a possible interpretation emphasizing the role of the associated currents.

\subsection{Event 1 - Alfv\'enic structure}
\label{sec:event1}

The first event we shall examine was
observed on November 6 during the time interval from 23:32:48 to 23:39:05.
The observed structure is rather typical and may be considered as representing quite large
group, if not the majority, that may be called Alfv\'enic structures. Plasma
parameters for this encounter are presented in Table~\ref{table:table_1}.

\begin{deluxetable*}{cccc}
\tablecaption{Major plasma parameters for the Alfv\'enic structure
\label{table:table_1}}
\tablewidth{0pt}
\tablehead{
\colhead{Parameter} & \colhead{Before encounter} & \colhead{Inside structure} & \colhead{After encounter} 
}
\startdata
magnetic field vector \si{nT} & [-75.8;-17.5;22.5] & [-46.1;27.5;-63.6] & [-74.4;18.8;23.0] \\ 
magnetic field magnitude \si{nT} & 81.0 & 83 & 80 \\ 
velocity vector \si{km/s} & [323.4;26.7;31.4] & [360.3;49.6;-74.6] & 
[330.6;56.3;33.9] \\ 
velocity magnitude \si{km/s} & 326 & 371 & 337 \\ 
ion plasma density \si{cm^{-3}} & 324 & 412 & 361 \\ 
ion temperature \si{eV} & 18 & 57.2 & 28.4 \\ 
ion beta  & 0.36 & 1.47 & 0.6 \\ 
ion inertial length c/$\omega_{pi}$ km & 12.6 & 11.2 & 12.0 \\ 
ion Larmor radius V$_{Ti}/\Omega_{i}$ km & 5.2 & 9.5 & 6.4
\enddata 
\end{deluxetable*}

\begin{figure*}
	\centering
    \includegraphics[width=\linewidth]{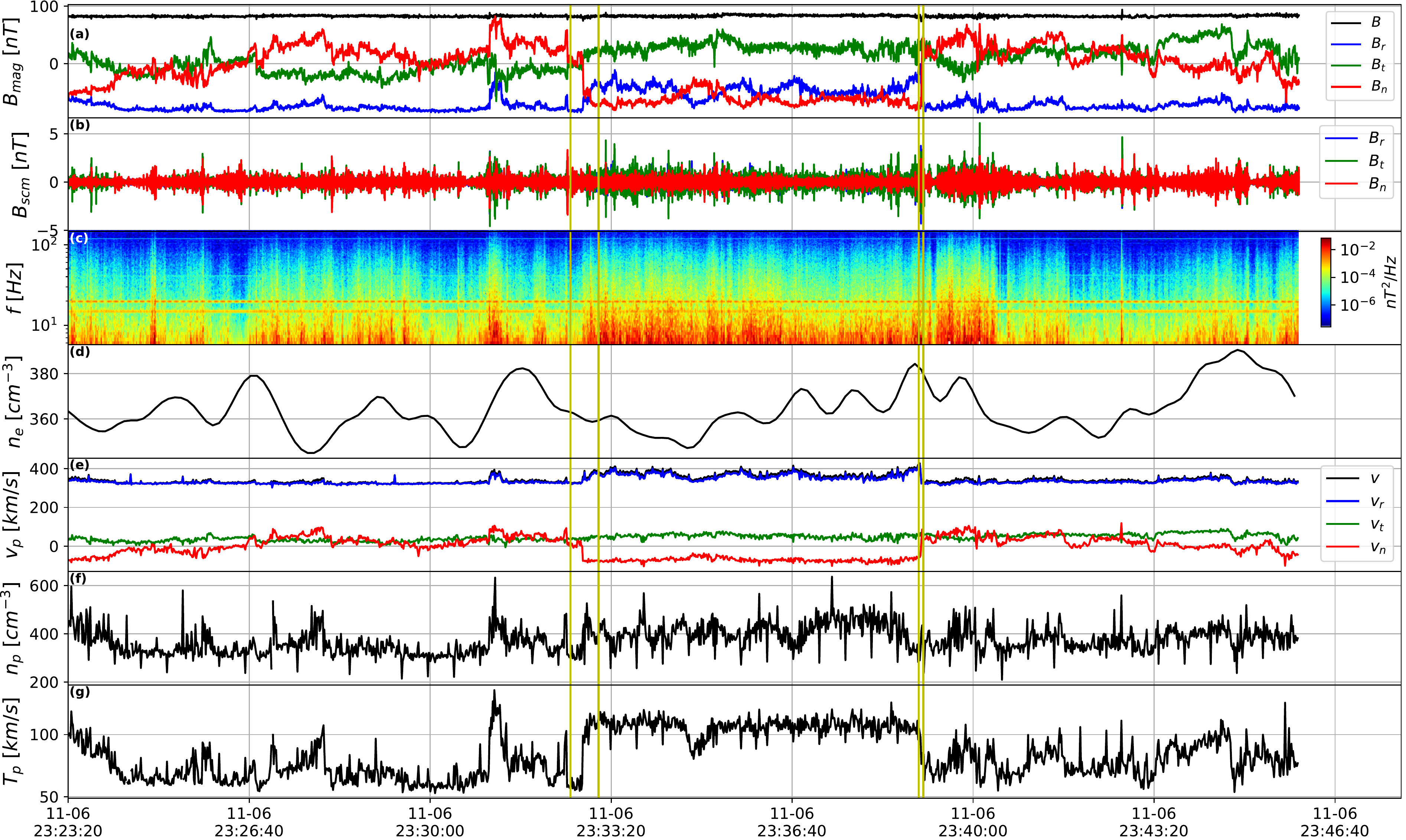} 
    \caption{The Alfv\'enic type magnetic structure detected on November 6, 2018: a) the magnetic field components in the RTN reference frame (the radial component is in blue, the tangential one in  green and the normal one in  red). The magnetic field magnitude is shown with the black curve. b) magnetic field fluctuations from the SCM instrument (the colors are the same as in panel~a. c) the magnetic field fluctuations dynamics spectrum. d) the averaged} electron plasma density from the QTN technique e) the plasma flow velocity from the SWEAP instrument measurements (the components colors are the same as in panel~a. f) the ion (proton) density from the SWEAP instrument measurements. g) the ion (proton) thermal velocity from the SWEAP instrument measurements.
    \label{fig:fig_2}
\end{figure*}

Figure~\ref{fig:fig_2}
represents the data of measurements of major plasma parameters during this
event. Panel~a shows magnetic field measurements by MAG instrument. The  data are expressed in  the RTN  frame (see Sec.~\ref{sec:data}): the subscript \textbf{r} stands for radial direction (blue lines), \textbf{n} for the normal one (red) and \textbf{t} for the tangential one  (green). The
total magnetic field is in black.

\bigskip

Figure~\ref{fig:fig_2} represents the parameters of the structure during an encounter on November 6 2018: the panel~a  shows the variations of the magnetic field in RTN reference frame as registered by MAG instrument. The panel~b shows the measurements of the three components of the magnetic field variations in the range from 5.7 Hz to 146 Hz registered by SCM instrument. The panel~c shows the spectra of the magnetic field obtained from SCM measurements in the same frequency range, here the horizontal lines correspond to interferences due to rotation of inertial wills ensuring pointing in the direction of the Sun. Panel~d presents an electron density as estimated from electric field measurements
using QTN technique. Panel e) shows three components of the proton flow
velocity evaluated from SWEAP distribution function measurements. The same
colors are used for velocity components as for the magnetic field, blue for
radial, green for tangential and red for normal (positive northward). Panel f) shows proton density estimated making use of SWEAP measurements, and panel g) the evaluation of the proton thermal velocity from the SWEAP instrument.
On Figure~\ref{fig:fig_2}, panel a) shows that the encounter begins from strong increase of the normal component of the south directed
magnetic field. It grows from several \si{nT} to more than \SI{50}{nT} and at the
same time the magnitude of the radial component decreases from about 80 \si{nT} to 60 \si{nT} and
than continues to decrease slowly inside the structure to about \SI{40}{nT}, remaining being negative all the time. The
normal southward component of the magnetic field becomes dominant that
results in the deviation of the magnetic field to about \ang{62} with
respect to magnetic field before encounter. At the same time the magnitude of the normal
(RTN) component of the velocity shown on the panel~d of the same Figure
increases from zero to about \SI{80}{km/s}, and the radial component
slightly increases also. The event lasts up to 23:39:04/05 when the magnetic
field returns back to the value close to its initial in 5 seconds, the same
does the velocity vector. It is worth noting that during all these
variations the magnitude of the magnetic field remains almost constant, as
it should be for Alfv\'enic structures. The second feature, almost synchronous
variation of the velocity and the magnetic field, with the precision of $
0.3$ sec for particle moments evaluated from measurements by SWEAP instrument
validates it also. Here it is necessary to mention that as an evaluation of
the moments of the ion distribution function needs 0.8 seconds, thus the
notion of synchronous means that two processes may be treated as those if
the time shift between FIELDS and SWEAP instruments measurements is less than 0.8 sec. The
two techniques of estimate of the plasma density, QTN and averaging over ion
distribution function are complementary. QTN determines with the high
precision a local electron density making use of the position of the peak in
the spectra corresponding to the zero of the dielectric permittivity \citep{Meyer2017,moncuquet20}. When the electron gyrofrequency is much
smaller than the plasma frequency (it is our case) this peak position as a
function of frequency is very close to the plasma frequency that allows one
to evaluate the plasma density with high precision. An important advantage
of this technique consists in its independence of evaluation on potential of
the satellite and of instrument itself. The weak point is related to the
need to averaging over several tens of points in order to evaluate the
position of the peak with the high precision. This leads to the smoothing of
the sharp density fluctuations. The determination of the density (and other
moments of the ion distribution function) making use of the Faraday cups
technique \citep{Kasper2016} is on the contrary dependent upon floating
potential of the satellite and of instrument itself and also on the angular
view of the instrument. From the other hand it allows to have more rapid
measurements that is quite important in the studies of the relatively small
scale fluctuations and sharp boundaries of structures. This explains the
difference of the density evaluation presented in the panels d) and f) of
Figure~\ref{fig:fig_2}. The thermal velocity shows up large and sharp increase on the leading edge and decrease on the trailing edge of the structure, the
variance inside is two times higher than outside that corresponds to
temperature four times higher. An estimate of the ion temperature and the
plasma ion beta (using density measurements by the SWEAP instrument) shows that
plasma ion beta outside the structure is about 0.36 before the encounter and
about 0.6 after. The temperature inside the structure is almost four times
higher, thus the boundary according to our data represents rather sharp
transition for ion $\beta $ from $\beta \simeq 0.36$ to $\beta $ about 1.5.
However the total dynamic pressure remains significantly larger.

It is worth noting the presence of important perturbations of the fields,
plasma density and flow velocity in the surrounding plasma. There is rather
intense wave activity just before the leading edge of the structure,
characterized by an important increase of the normal component of the
magnetic field from the almost negligible value to almost \SI{90}{nT} accompanied
by an increase in the density, velocity and ion temperature with
rather rapid pre-return and then slower return to previous values, where the
magnetic field magnitude is also slightly perturbed.

\begin{figure*}
	\centering
    \includegraphics[width=\linewidth]{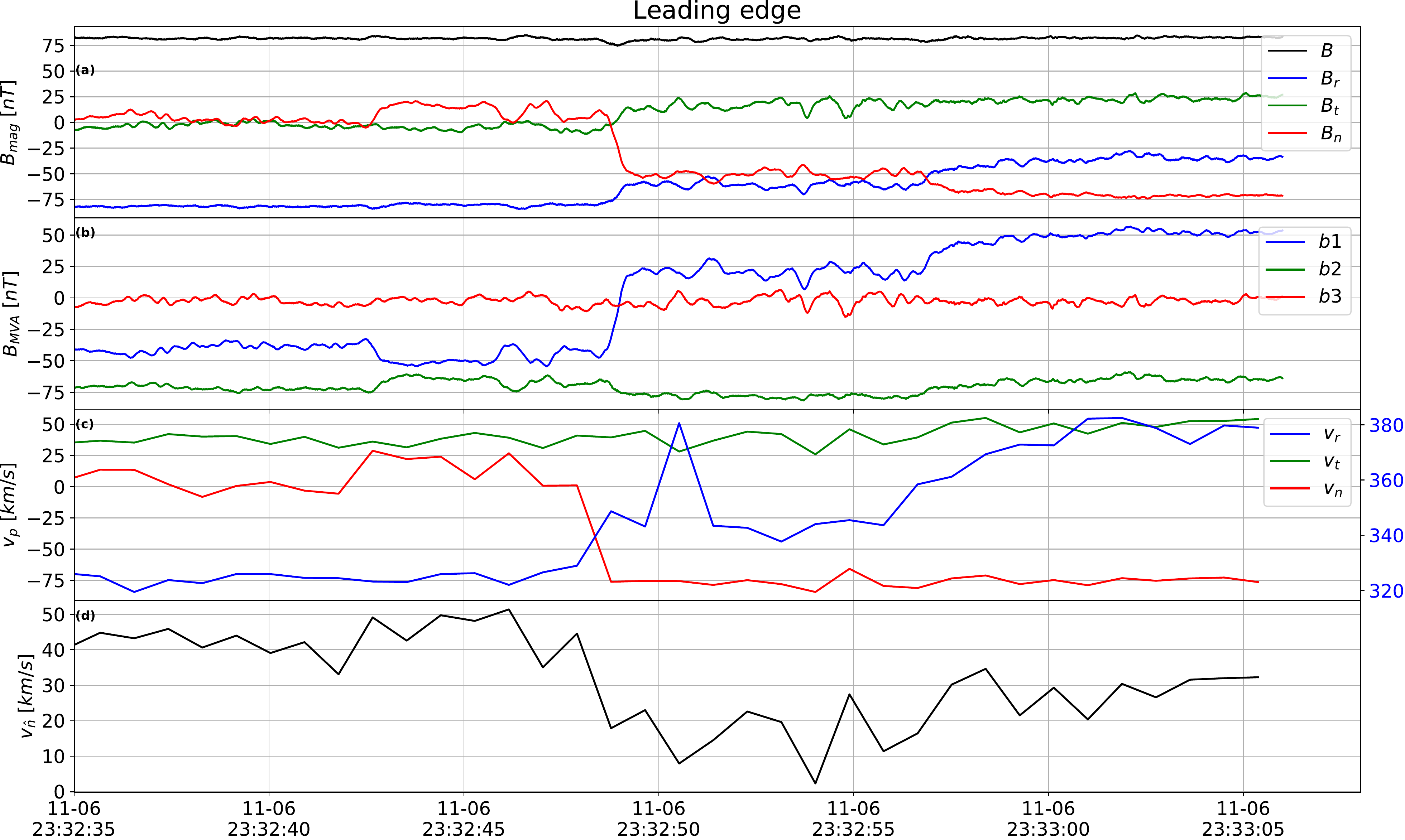} 
    \caption{The magnetic field (panel~a: in the RTN frame; panel~b: in the MVA frame) and plasma bulk velocity (panel~c: in the RTN frame and panel~d: the normal to the boundary component) on the leading edge the structure shown in Figure~\ref{fig:fig_2}. The components colors in RTN frame are similar to the ones in Figure~\ref{fig:fig_2}. In MVA frame the colors are used
differently than on other Figures, blue to mark the component corresponding
to the largest eigenvalue, i.e. corresponding to largest variation of the
field, green to the component corresponding to intermediate value, and red
to notify the one related to the smallest eigenvalue, On panel c the largest velocity being strongly different from the others is scaled on the right vertical line.}
    \label{fig:fig_3}
\end{figure*}

Now we shall examine in more detail the boundaries of the structure. In order to do it we use a technique called Minimum Variance Analysis (MVA). The MVA technique for the current sheets described in detail in \citep{Sonnerup1998}, determines three eigenvalues
and corresponding eigen-vectors. The largest eigenvalue shows the direction
of the largest change of the magnetic field, and transformation of the field
onto corresponding reference frame provides the magnitude of total magnetic
field change in the corresponding direction. The minimum variance vector
determines the direction of spatial variation of the field. Since the
Amper's law states that the current flows to the direction perpendicular to
both maximum jump of the field and the direction of its variation the
eigenvector corresponding to intermediate eigenvalue determines the
direction of the current flow. Since the SWEAP measurements provide the flow
velocity vector, its projection to the direction of the normal to the
current sheet and time of the boundary crossing allow one to evaluate the
characteristic spatial scale of the current sheet.
Figure~\ref{fig:fig_3} shows the boundary transition at the leading edge of the structure and the
results of the Minimum Variance Analysis of the MAG and SWEAP data. It
shows the variations in the RTN frame and in the reference frame
defined by MVA for the magnetic field and velocity.

\bigskip 

Upper panel~a of the Figure shows the variation of the magnetic field in
the RTN frame as a zoom of Figure 2 on the leading edge of the structure. The transition occurs at approximately 23:32:47/48 and
in about a second the magnitude of the normal component of the magnetic
field decreases from small positive value of about \SI{10}{nT} up to \SI{-50}{nT}
while the magnitude of the radial component at the same time decreases from about \SI{80}{nT} to
about \SI{60}{nT}. Two components become comparable and during further slow
evolution the normal component continues to increase and the magnitude of
radial one is decreasing and the normal component becomes dominant with the
field magnitude about \SI{70}{nT}. The magnitude of the magnetic field slightly
increases also from \SI{81}{nT} outside to \SI{84}{nT} inside the structure. Panel~b
shows variations of the magnetic field vectors corresponding to three
eigenvalues found by means of the MVA analysis. Here the colors are used
differently than on other Figures, blue to mark the component corresponding
to the largest eigenvalue, i.e. corresponding to largest variation of the
field, green to the component corresponding to intermediate value, and red
to notify the one related to the smallest eigenvalue. \textbf{This} last
determines also the direction of the normal to the transition considered as
discontinuity. Panel~c shows synchronously occurring evolution of the
velocity vectors. Here the radial velocity is shifted down to $300 \si{km/s}$,
to show the variations having comparable magnitudes. The radial velocity
changes from $323 \si{km/s}$ outside to $360 \si{km/s}$ inside. The normal component
of the velocity drastically changes from $22 \si{km/s}$ in the south-north
direction to $75 \si{km/s}$ in the opposite north-south. The total velocity
increases and the difference corresponds approximately to Alfv\'en velocity.
Taking into account that the magnetic field magnitude is about $81-83$~\si{nT},
and the plasma density about \SI{400}{cm^{-3}} one can find that Alfv\'en velocity
is about $100 \si{km/s}$. The SWEAP instrument data here as mentioned before is
shifted to 0.3 seconds. For this event we have evaluated the direction of
the normal to the boundary for the velocity vector also and it is very close
to the one found for the magnetic field (within \ang{9} of difference). Since the transition for the velocity vectors coincides with the transition for the magnetic field MVA analysis for them is very similar to that of for magnetic field.
Panel~d shows the variation of the normal to the boundary component of the
velocity that varies from about $40 \si{km/s}$ outside the structure to
approximately $10 \si{km/s}$ inside but then grows to larger values. The normal
vector to the structure at the leading edge is close to tangential axes with
small component along vertical, while the radial component is negligible.
The normal component of the magnetic field to this boundary (panel~b on Figure~\ref{fig:fig_3}) is equal to $B_n =$\SI[separate-uncertainty = true]{-2.5 \pm 3.2}{nT}, i.e., it is practically negligible.

\begin{figure*}
	\centering
    \includegraphics[width=\linewidth]{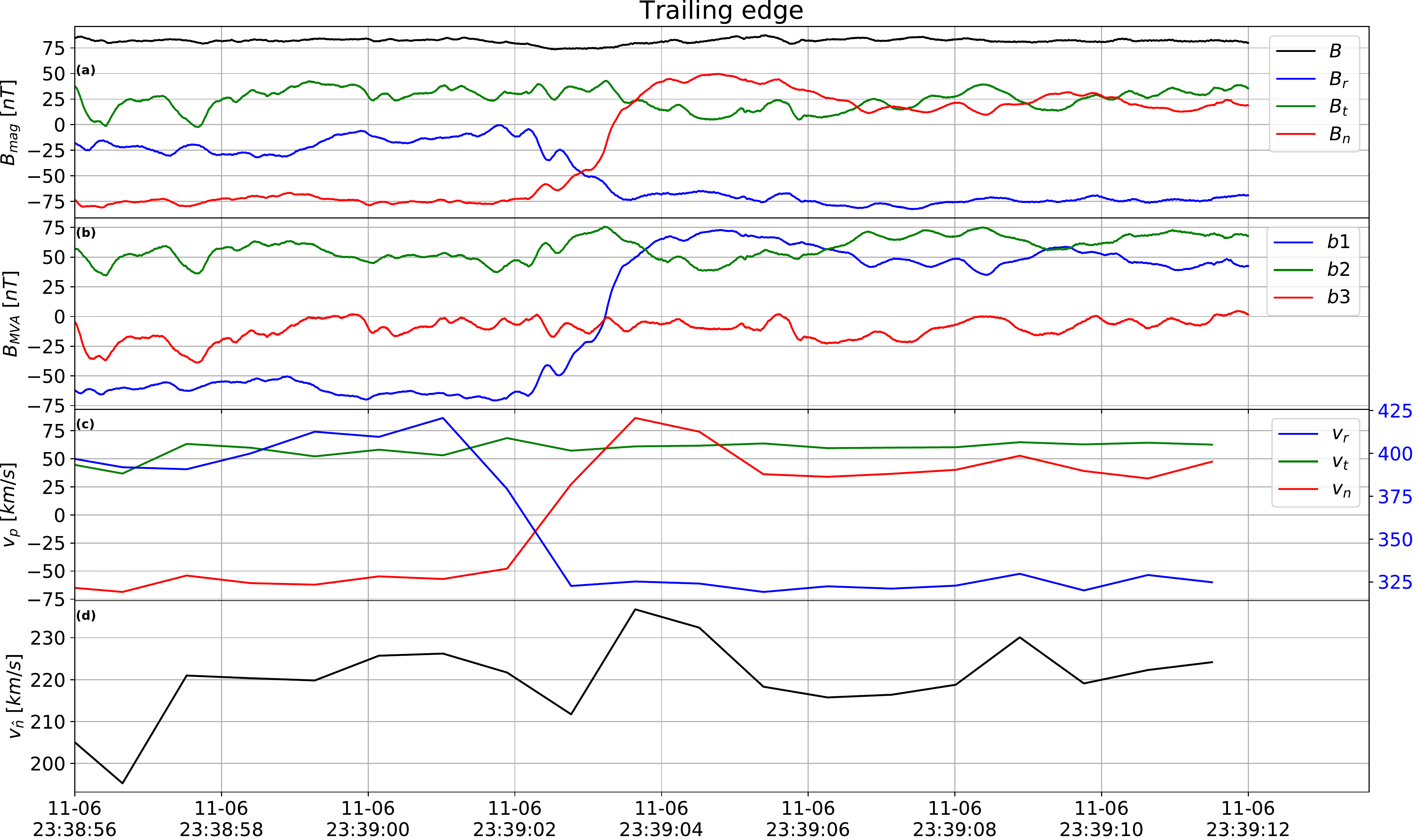} 
    \caption{The magnetic field and plasma bulk velocity on the trailing edge of the structure shown in Figure~\ref{fig:fig_1}. The format is the same as in Figure~\ref{fig:fig_3}.}
    \label{fig:fig_4}
\end{figure*}

Figure~\ref{fig:fig_4} shows variations of the fields and velocities across the trailing
edge boundary. The crossing of the trailing edge takes about 2 seconds, a
little longer than that of the leading edge, and the normal velocity is much
larger than that of the leading edge, thus the thickness of the boundary is
also significantly larger. Panel~a presents variations of the components of
the magnetic field in the RTN frame. The radial component decreases from the
value of about \SI{-25}{nT} to \SI{-74}{nT}, at the same time the normal component
returns from the large negative value of about \SI{-75}{nT} to a positive value
of about \SI{50}{nT}. The total magnetic field slightly slowly decreases also.
The variations of the magnetic field in the MVA
frame are presented on panel~b. The evolution of two major components of
eigenvectors consists in slow rotation of the vector in clockwise direction
with some wave activity around departure and arrival points. Panel~c shows
variations of the velocity vector, radial component is downshifted to
300 \si{km/s}, the tangential component remains almost unchanged, the radial
velocity decreases from 420 to \SI{323}{km/s}. The normal component of the velocity changes from \SI{60}{km/s} towards south to \SI{80}{km/s} towards the north. The total velocity decreases
also from \SI{428}{km/s} to \SI{328}{km/s}. The normal vector to the boundary 
is
found to be $\mathbf{N}=[0.48;0.8;0.35]$, the largest
component is along the tangential axes, but radial and normal components are
also quite significant. Normal component of the magnetic field is $B_n =$\SI[separate-uncertainty = true]{-13.3 \pm 5.8}{nT}, still significantly smaller than the
magnetic field magnitude. The variations of the magnetic field and velocity
occur synchronously within precision corresponding to characteristic
sampling rate of particle measurements. Panel~d shows the variations of the
normal velocity across the boundary. It varies in the range $200-235$ $%
 \si{km/s} $.

An important characteristic of the Alfv\'enic fluctuations consists in linear
relation between the perturbations of the velocity and the magnetic field.
As it was already noted by \citet{kasper19} in
majority of cases they vary simultaneously that validates the hypothesis
that the fluctuations are Alfv\'enic. In MHD approximation the relation
written above implies: 
\begin{equation*}
\mathbf{\delta V}=c\mathbf{V}_{A}\frac{\delta \mathbf{B}}{B}
\end{equation*}
here c is a constant.
\begin{figure}
	\centering
    \includegraphics[width=\linewidth]{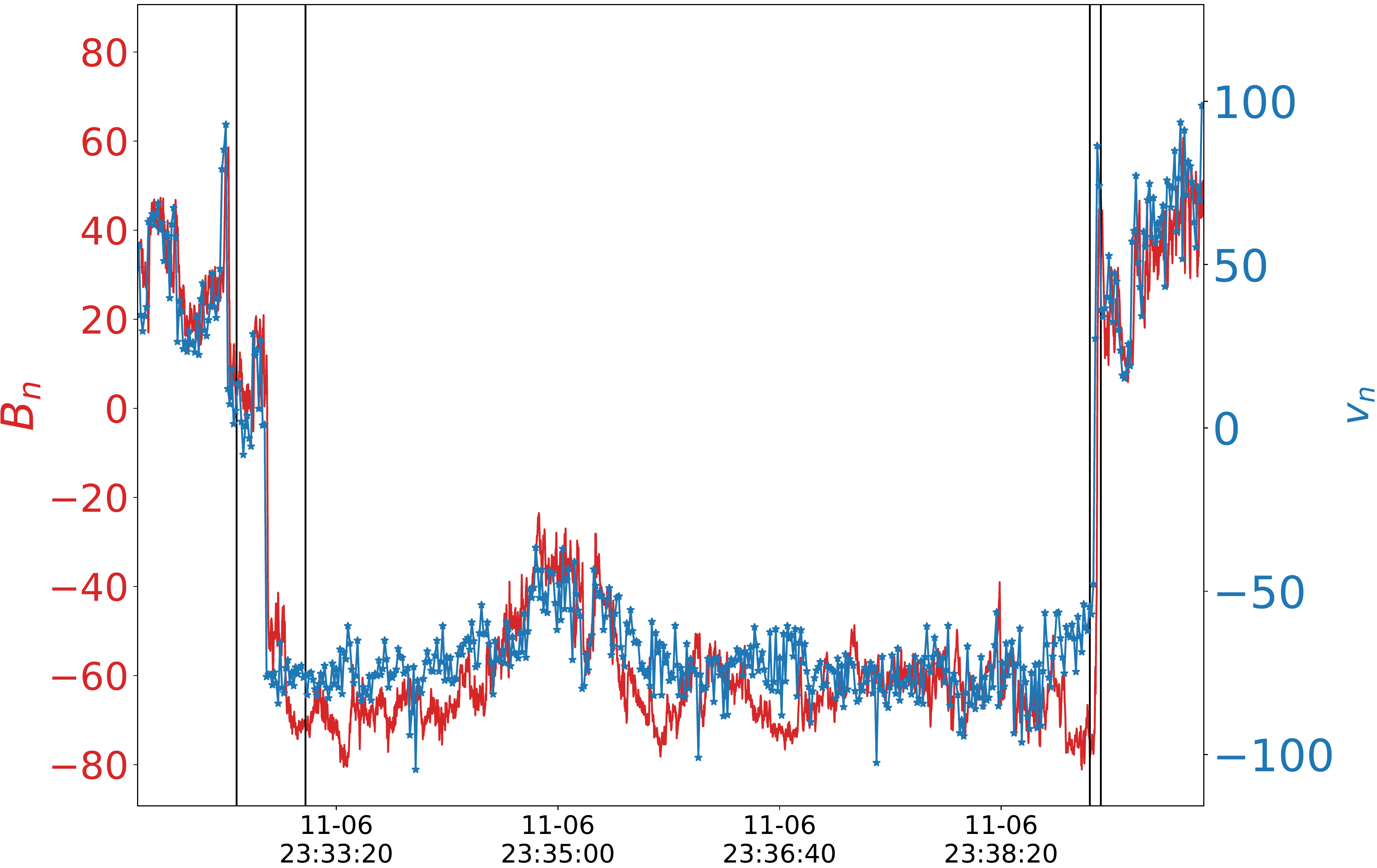} 
    \caption{The normal components of the magnetic field (red) and plasma bulk velocity (blue) during crossing of the structure (Figure~\ref{fig:fig_2}), showing linear correlation. The leading and trailing boundaries are marked by the vertical lines. The variations on the boundaries are synchronous and fit the linear relation. }
    \label{fig:fig_5}
\end{figure}

On Figure~\ref{fig:fig_5} we present the variations of the normal components of the
magnetic field and velocity that undergo the largest variations. Vertical
lines show the areas of the major transition for the leading and trailing
edges. The jump of the velocity on the leading edge boundary is about $%
\mathbf{\delta V}=\SI{100}{km/s}$, corresponding jump of the magnetic field is $%
\frac{\boldsymbol{\delta }\overrightarrow{B}}{\boldsymbol{B}}\mid 
\boldsymbol{V}_{A}\mid =\SI{93}{nT}$, thus the coefficient $c$ is about 1.08,
similar values for the trailing edge $\mathbf{\delta V}=\SI{141}{km/s} ,$ $\frac{%
\boldsymbol{\delta }\overrightarrow{B}}{\boldsymbol{B}}\mid \boldsymbol{V}%
_{A}\mid =\SI{125}{nT}$, give the value 1.12. These numbers are in the range of
error bars on the actual level of instruments calibrations, thus the
coefficient is about 1.
\begin{figure*}
	\centering
    \includegraphics[width=\linewidth]{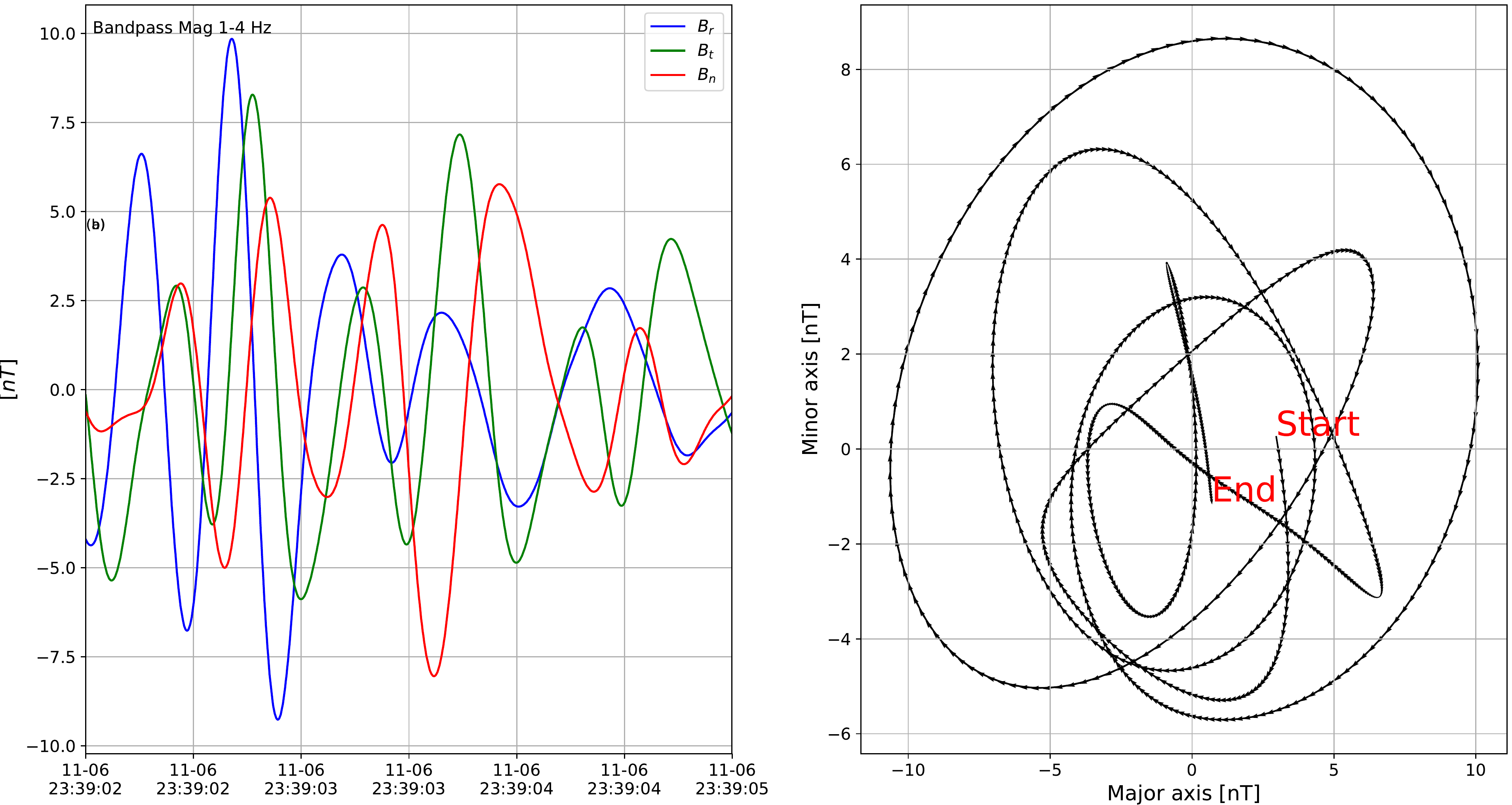} 
    \caption{a) Waveforms of the magnetic field fluctuations in the RTN frame filtered in 1-4~\si{Hz} frequency band the 3 s time interval on the trailing edge of the structure shown in Figure~\ref{fig:fig_2}. b) Hodograph of the magnetic field (in the plane transverse to the wave normal) showing close to the circular polarization of the observed wave.}
    \label{fig:fig_6}
\end{figure*}

The wave activity around the leading edge boundary is relatively weak. But
it is quite intense around the trailing edge. Figure~\ref{fig:fig_6} shows its
manifestations around it. On the left panel~a the three components of
magnetic field variations registered by MAG instrument in frequency range
\SIrange{1}{4}{Hz} are presented for short time interval around the boundary from
23:39:02 to 23:39:05, the colors are similar to other Figures. The
amplitudes of oscillations may become as large as \SI{10}{nT}. The wave magnetic
field rotates as shown on panel~b where the hodograph of vectors
corresponding to two largest eigenvalues are shown. We determine the $k$-vector of the wave and found that it makes an angle of 
\ang{60.5} with the boundary normal. This provides a strong indication
that the wave mode corresponds to surface wave \citep{Hollweg1982}. This wave activity is manifested in strong increase of the local wave energy flux. 

\begin{figure*}
	\centering
    \includegraphics[width=\linewidth]{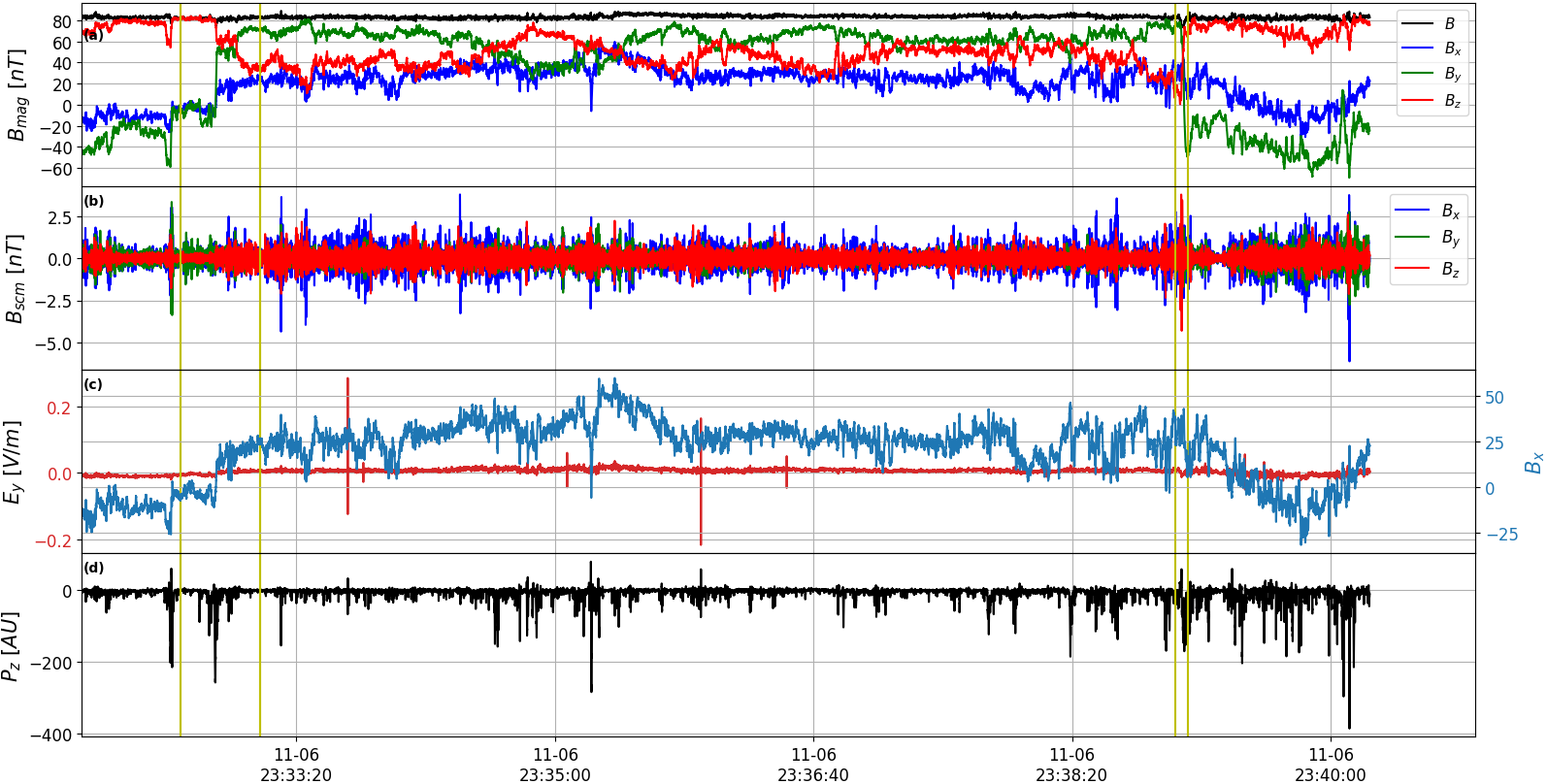} 
    \caption{Poynting flux evaluation. a) Magnetic field measurements by MAG instrument during the structure crossing in satellite xyz reference frame. Since there are only two components of the electric field measured on board PSP x and y, for this particular Figure the data is presented in satellite reference frame, contrary to other Figures. The components are colored x in blue, y in green and z in red. b) Magnetic field fluctuations registered by the SCM instrument, colors are similar to panel a. c) y-component of the electric field registered by FIELDS suite, in red, and x-component of the magnetic field by MAG instrument in blue. These fields components are used together with the x-component of the electric field and y-component of the magnetic field for evaluation of the z-component of the Poynting flux. d) z-component of the Poynting flux during time interval including Alfv\'enic structure crossing. Vertical lines mark the intervals around structure leading and trailing edges.}
    \label{fig:fig_7}
\end{figure*}

Figure~\ref{fig:fig_7} represents an evaluation of the Poynting flux estimated for the time interval covering crossing of the structure. On upper panel~a the magnetic field variations are shown to determine the
timing and positioning of fields and the Poynting flux with respect to
boundaries. The magnetic field fluctuations registered by SCM instrument are
shown with the spiky bursts of large amplitude waves in the range of several
Hz on the trailing edge boundary of the structure as well as inside it. It is worth reminding that the wave activity around leading edge boundary is significantly weaker, and the Poynting flux is also much smaller. Panel~c
represents an evaluation of the z-component of the Poynting flux calculated
in the plasma reference frame. The flux calculated in satellite reference frame might contain artificial enhancements and attenuations caused by variations of plasma speed around satellite. Velocity variations create artificial variations of the electric field due to V cross B induced electric field, that in its turn results in artificial variations of the Poynting flux. In plasma reference frame this velocity pollution effect is excluded. The measurements of two components of the electric and magnetic fields allow one to evaluate the $z-$component of the
Poynting flux. The procedure of calibration of electric field making use of
the magnetic field measurements is described in detail by \citet{Mozer2020}. It is worth mentioning that z-component in the
satellite reference frame is very close to radial direction. The Poynting
flux in overwhelming majority of events where it is significant is directed
towards interplanetary space. However, there exists some quite short
intervals when it is directed oppositely, towards the Sun. The intervals of
sharp increase of the Poynting flux coincide with the bursts of the wave
activity, as it is quite intense around the trailing edge boundary the Poynting flux strongly increases there. 

\begin{figure*}
	\centering
    \includegraphics[width=\linewidth]{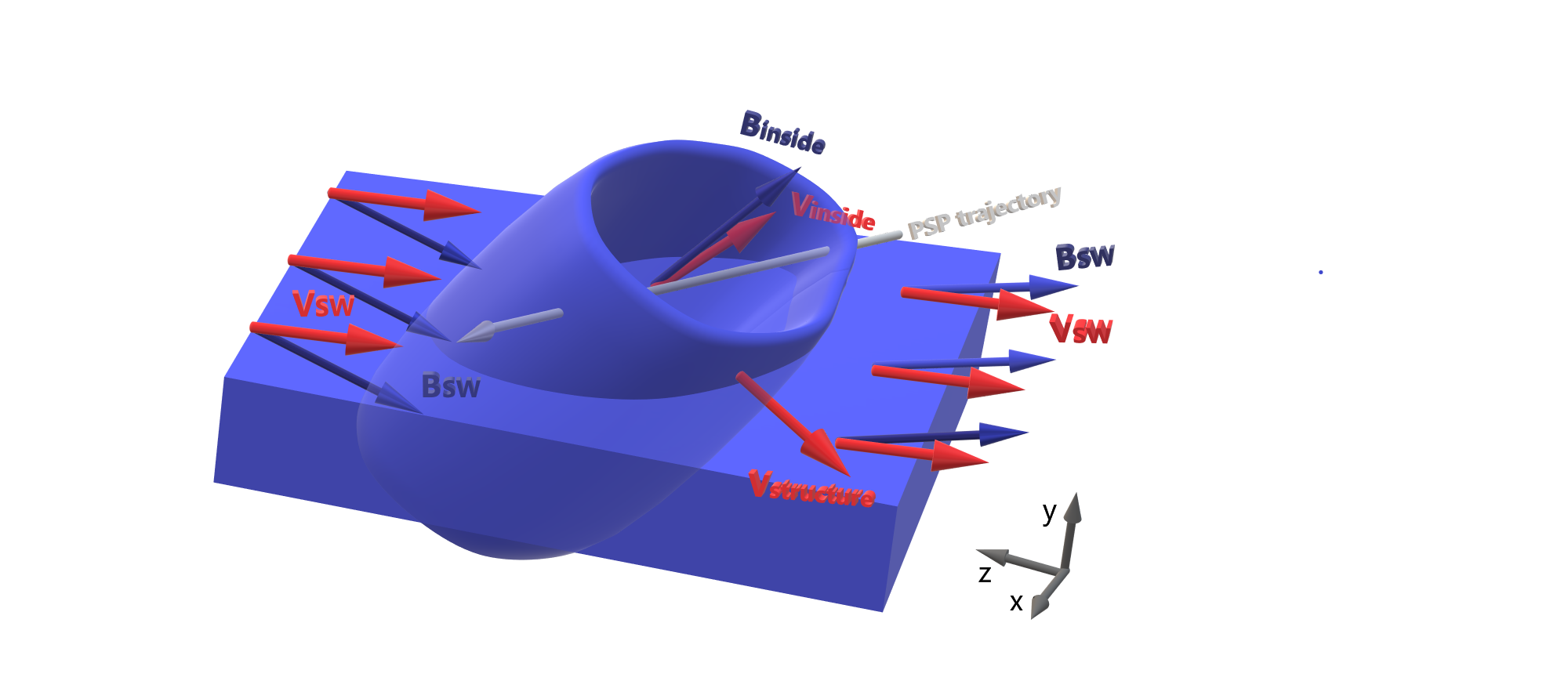} 
    \caption{The schematic illustration of the magnetic structure shown in Figure~\ref{fig:fig_2}. The solar wind plasma bulk velocity is shown by the red vectors $\mathbf{V}_{SW}$ (for the background solar wind) and $\mathbf{V}_{inside}$ (inside the structure); the magnetic field is shown by the blue vectors $\mathbf{B}_{SW}$  and $\mathbf{B}_{inside}$, respectively. The solar wind bulk flow velocity ahead and behind of the encounter is about the same while the magnetic field vectors have slightly different directions indicating the current system on the boundary of the structure. The velocity of the plasma flow inside the structure has two components: the flow along the field lines parallel to the tube axes $\mathbf{V}_{inside}$ (along the magnetic field in the structure), and as the component related to the structure motion $\mathbf{V}_{structure}$. Thus the total bulk flow velocity of the plasma inside the structure is $\mathbf{V}_{flow}$ = $\mathbf{V}_{inside}$ + $\mathbf{V}_{structure}$.  The gray line illustrates the trajectory of the satellite crossing the structure.}
    \label{fig:fig_8}
\end{figure*}

The data set presented for this event admits quite natural interpretation of
the observations in terms of the crossing of magnetic tube as presented by the sketch drawing on Figure~\ref{fig:fig_8}. The
characteristic spatial scales of plasma for this event are ion Larmor radius
and the ion inertial length. The ion Larmor radius varies in the range 5 -
9.5 km, the ion inertial length 11 - 12 km (see Table~\ref{table:table_1}). These
characteristic scales are significantly smaller than the scales of
variations of the parameters of macroscopic plasma motions in the whole
region except the boundaries. This implies that the macroscopic plasma flow
may be treated in terms of "frozen in motions" and in MHD approximation. It
means that the plasma moves along the field lines of the magnetic field. Our
data testify that the satellite traverses the region where plasma parameters
and fields strongly differ of those outside the structure. In order to
better understand physical processes related with this magnetic tube the
motion of the plasma may be separated into two parts, the motion of the
plasma flow along the field lines inside and outside the tube, that we shall notify as $\mathbf{V}_{par}$ and the motion of the magnetic tube itself perpendicularly to its axes notified correspondingly as $\mathbf{V}_{\perp }$.

The estimates of the velocities of the plasma flow and magnetic field tubes
for the structure we examine give following results. Before the encounter
the flow moves along the field lines with the velocity $297,7 \si{km/s}$, and
magnetic field lines (tubes) have relative velocity with respect to
satellite in the orthogonal direction equal to 
\begin{equation*}
\mathbf{V}_{1\perp }=[46.1;-35.5;-111.4]~\si{km/s} .
\end{equation*}%
There is no way to define the tube's velocity along their axes, thus we
attribute it to the plasma flow motion. One can see that the tube lines are
directed mostly radially from the Sun and they move preferentially in the
direction from north to south with some deviation towards tangential
direction. After crossing the leading edge of the structure and entering
inside it the magnetic field drastically changes its direction, and the
average velocity vector also changes. Making estimate of the velocity of
plasma flow along the field lines inside the structure one can find it to be
equal to $129.2 \si{km/s}$, and the velocity of the tube in the orthogonal
direction is presented by the vector: $\mathbf{V}_{2\perp }%
=[287.6;91.6;-172.4]$ with perpendicular velocity equal to $V_{2\perp
}=347.6~\si{km/s} $. \citep{kasper19} called such
structures "jets", because the total measured velocity with respect to
external observer, in our case satellite, inside the structure is indeed
larger than the total velocity of the flow and field lines outside the
structure. 

The angle between internal and external tubes before encounter is
\ang{75.5}, the angle with the tubes after crossing the structure is \ang{67.6}. The magnetic field tube of the structure moves mostly in the
direction of the vertical-radial plane with the angle of elevation about \ang{51}, and its motion is mainly in the same plane in radial outward -
vertical north-south direction. The relative velocity of the combined motion
of the tube and plasma flow along the tube with respect to external flow
before is about \SI{113.5}{km/s}, which is close to the local Alfv\'en speed.

After crossing the trailing edge boundary the direction of the tubes changes
drastically again returning to direction close to initial. The flow velocity
parallel to the local magnetic field becomes equal to $V_{par}=284~\si{km/s} $,
and perpendicular velocity of the field lines is presented by the vector $\mathbf{V}_{\perp}=[209.4;211.5;-17.5]~\si{km/s}$.

Our interpretation of the structure as magnetic tube that one may imagine to
be cylindrical with circular or elliptical cross section provides an idea
about characteristic transverse scale of the structure. Normal component
of the velocity along the leading edge of the structure varies from $
V_{lead}$ \SIrange{10}{40}{km/s} and during crossing of the trailing edge $%
V_{trail}$ \SIrange{200}{220}{km/s}. The total duration of the satellite journey
inside the structure is approximately equal to $\delta t\simeq 7$ minutes.
Assuming that the component of the normal velocity along the satellite
trajectory varies from \SIrange{20}{200}{km/s} one can evaluate the total
distance along the chord to be equal to 
\begin{equation*}
L\simeq \frac{(V_{lead}+V_{trail})}{2}t\simeq 5\times 10^{4}~\si{km}.
\end{equation*}

Figure~\ref{fig:fig_8} illustrates the geometry of the local configuration of the magnetic
tube. It should be pointed out very important feature of our observations
that consists in the difference of the direction of the magnetic field
before and after the structure. The magnetic fields before and after the
structure make an angle of approximately \ang{30} that corresponds to
rather important shear of the magnetic fields, thus the structure as a whole
represent the current sheet. We shall discuss it in the last section in more
detail.


\subsection{Event 2 - Compressional structure} 
\label{sec:event2}

This event was registered at November 4 from 17:05:34 to 17:06:50. Its
crossing took shorter time interval than previous structures. This structure
represents another group of "switchbacks" that has different characteristics
than previous event. The major difference consists in quite strong variation
of magnitude of the magnetic field inside the structure. In average it is significantly smaller
inside the structure than outside, and at the same time the ion density
according to ion distribution measurements by SWEAP decreases also. The ion
temperature simultaneously with the decrease of magnetic field magnitude
increases. Moreover, the magnetic field undergoes quite strong variations
Inside the structure. The variations of the normal and radial components are
up to tens of percent. According to these observations, namely, correlation
of magnetic field magnitude and density decreases, in terms of the MHD waves
classification the structure should have belonged to fast magnetosonic
rather than Alfv\'enic wave mode. However, we shall demonstrate that there are
some important characteristics that are similar to Alfv\'enic structures.

The basic plasma parameters for this structure are listed in Table~\ref{table:table_2}.

\begin{deluxetable*}{cccc}
\tablecaption{Basic plasma parameters for compressional structure
\label{table:table_2}}
\tablewidth{0pt}
\tablehead{
\colhead{Parameter} & \colhead{Before encounter} & \colhead{Inside structure} & \colhead{After encounter} 
}
\startdata
magnetic field vector \si{nT} & [-68.6;16.6;-1.6] & [26.9;42.3;21.7] & 
[-65.5;-13.1;8.5] \\ 
magnetic field magnitude \si{nT} & 70.6 & 54.6 & 67.3 \\ 
velocity vector  \si{km/s} & [301.3;42.8;15.7] & [402.0;68.3;27.1] & 
[293.4;16.1;28.9] \\ 
velocity magnitude  \si{km/s} & 304.8 & 408.6 & 295.2 \\ 
ion plasma density \si{cm^{-3}} & 311 & 282 & 302 \\ 
ion temperature \si{eV} & 21.5 & 46.5 & 18.3 \\  
ion beta  & 0.54 & 1.8 & 0.49 \\ 
ion inertial length c$/\omega_{pi}$ \si{km} & 12.9 & 13.6 & 13.1 \\ 
ion Larmor radius V$_{Ti}/\Omega_{i}$ \si{km} & 6.6 & 12.5 & 6.4
\enddata 
\end{deluxetable*}

\begin{figure*}
	\centering
    \includegraphics[width=\linewidth]{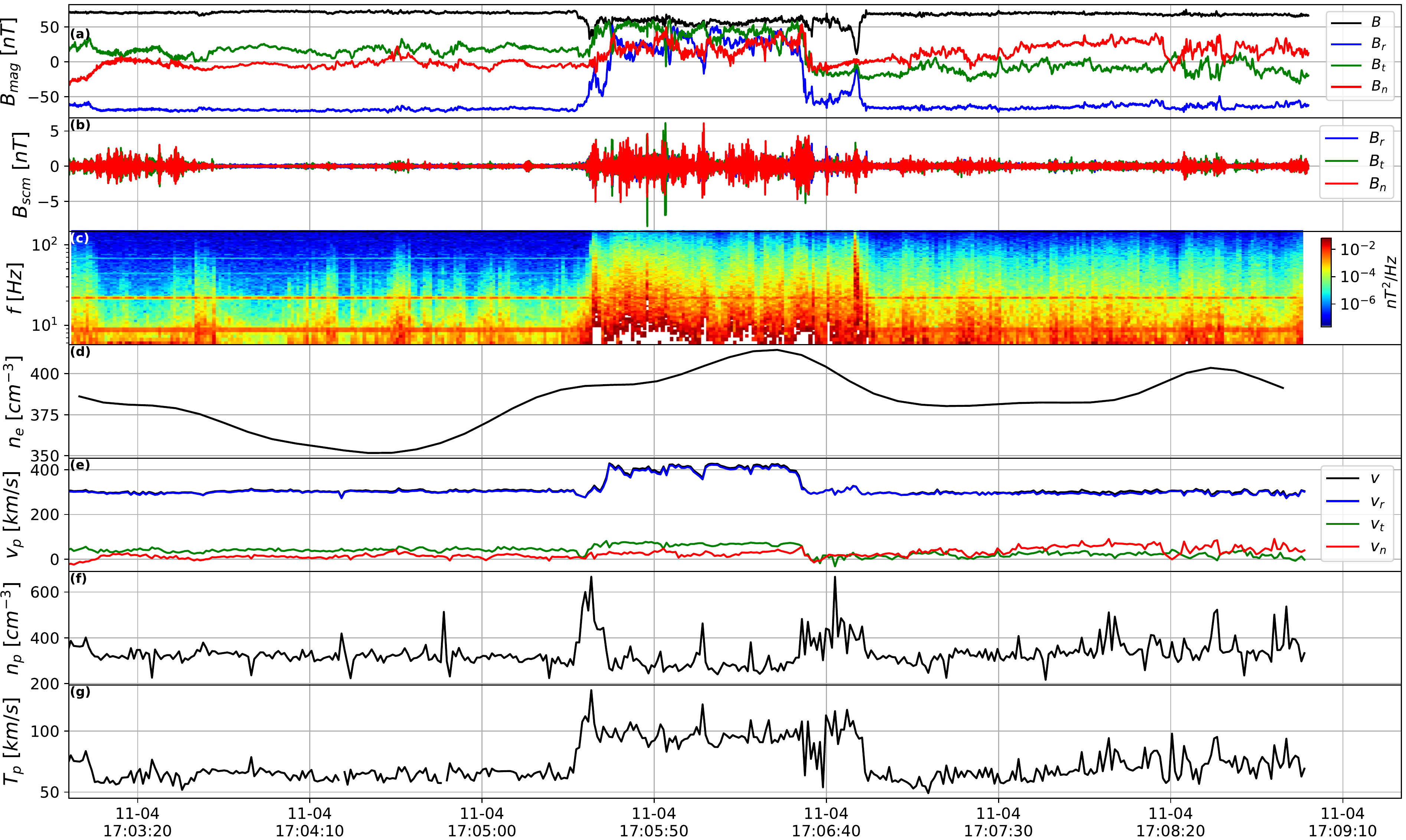} 
    \caption{The compressional magnetic structure in the same format as in Figure~\ref{fig:fig_2}.}
    \label{fig:fig_9}
\end{figure*}

Figure~\ref{fig:fig_9} represents the parameters of the structure: panel~(a) shows the three components of the magnetic field. The structure starts with the sharp decrease of the magnetic field magnitude from \SI{70}{nT} to \SI{30}{nT} and the simultaneous change of the radial component from about \SI{-70}{nT} to about \SI{-6}{nT}, followed by the partial rebound
to about \SI{-45}{nT} and then the major jump to \SI{52}{nT}. At the same time the
tangential component increases from about \SI{10}{nT} to \SI{60}{nT}, and the normal
component changes its sign and jumps from \SI{-10}{nT} to about \SI{40}{nT}. These
variations happen in about 10~s interval from 17:05:26 to 17:05:37. At
17:06:25 the magnetic field begins its return, the major jump takes place
from 17:06:26 to 17:06:36. An additional jump of the magnetic field occurs
at about 17:06:46 when the magnitude of the radial component of the magnetic
field decreases to negative values of several \si{nT}, and the total field
becomes small of the same order, and, reaching a minimum the field quickly
returns to quasistationary outside values. Panel~(b) shows waveform of
magnetic fluctuations recorded by the SCM instrument. They are very intense
with the dominant component in the normal direction and amplitudes as large
as several \si{nT} (up to 8~nT). Most intense oscillations are in frequency
range from 1-2 Hz to about 10 Hz. Bouncing of
the DC magnetic field just before and a little later after encounter are
filled in by very intense magnetic oscillations. Panel~(c) shows
spectra of magnetic fluctuations obtained from the SCM instrument measurements.
The magnetic fluctuations frequency range goes up to  \SI{100}{Hz}. The wave activity is
quite intense around both the boundaries. Panel~(d) shows electron density estimated by means of the QTN technique. 
In our study we use density,
bulk flow and thermal velocity measured by the SWEAP instrument as they are better
adapted for the studies of sharp boundaries. Panel~(e) provides an evaluation
of the thermal velocity of ions, and the strong increase of it inside the
structure results in a strong growth of the plasma ion beta from $\beta
\simeq 0.55$ outside the structure to $\beta \simeq 1.8$ inside, and then
rebounds to $0.49$ after encounter. A possible role of sharp ion plasma
beta variations for the equilibrium of the structure will be discussed
later. It is worth noting that the dominant parameter in comparison of
pressure balance remains the dynamic pressure of the plasma flow. Another
important features are the density enhancement on the leading and trailing edge (begins inside the structure) presumably related
to the plasma drag on the boundaries.

\begin{figure*}
	\centering
    \includegraphics[width=\linewidth]{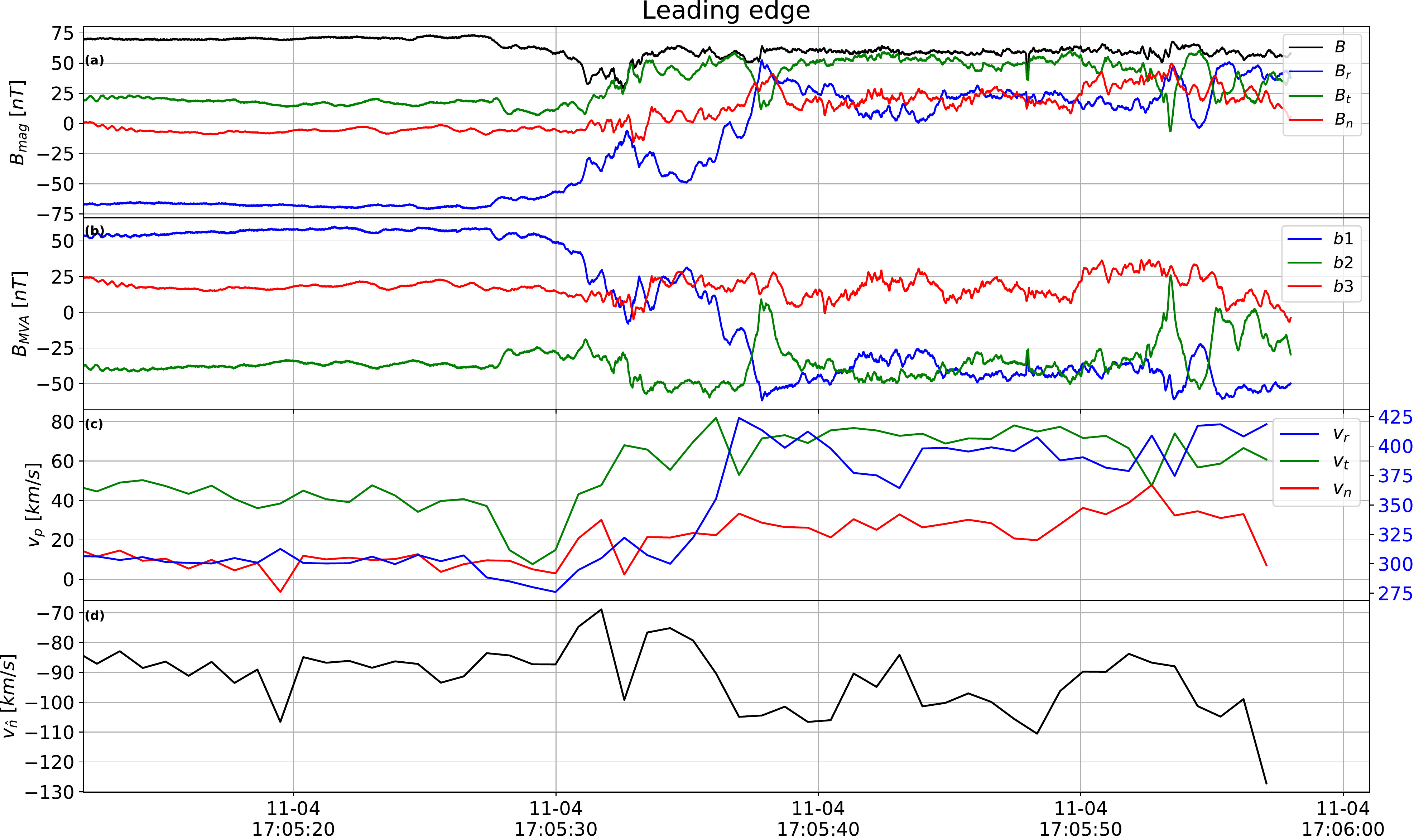} 
    \caption{The magnetic field and plasma bulk flow velocity on the leading edge of the compressional structure shown in Figure~\ref{fig:fig_9}.}
    \label{fig:fig_10}
\end{figure*}

\begin{figure*}
	\centering
    \includegraphics[width=\linewidth]{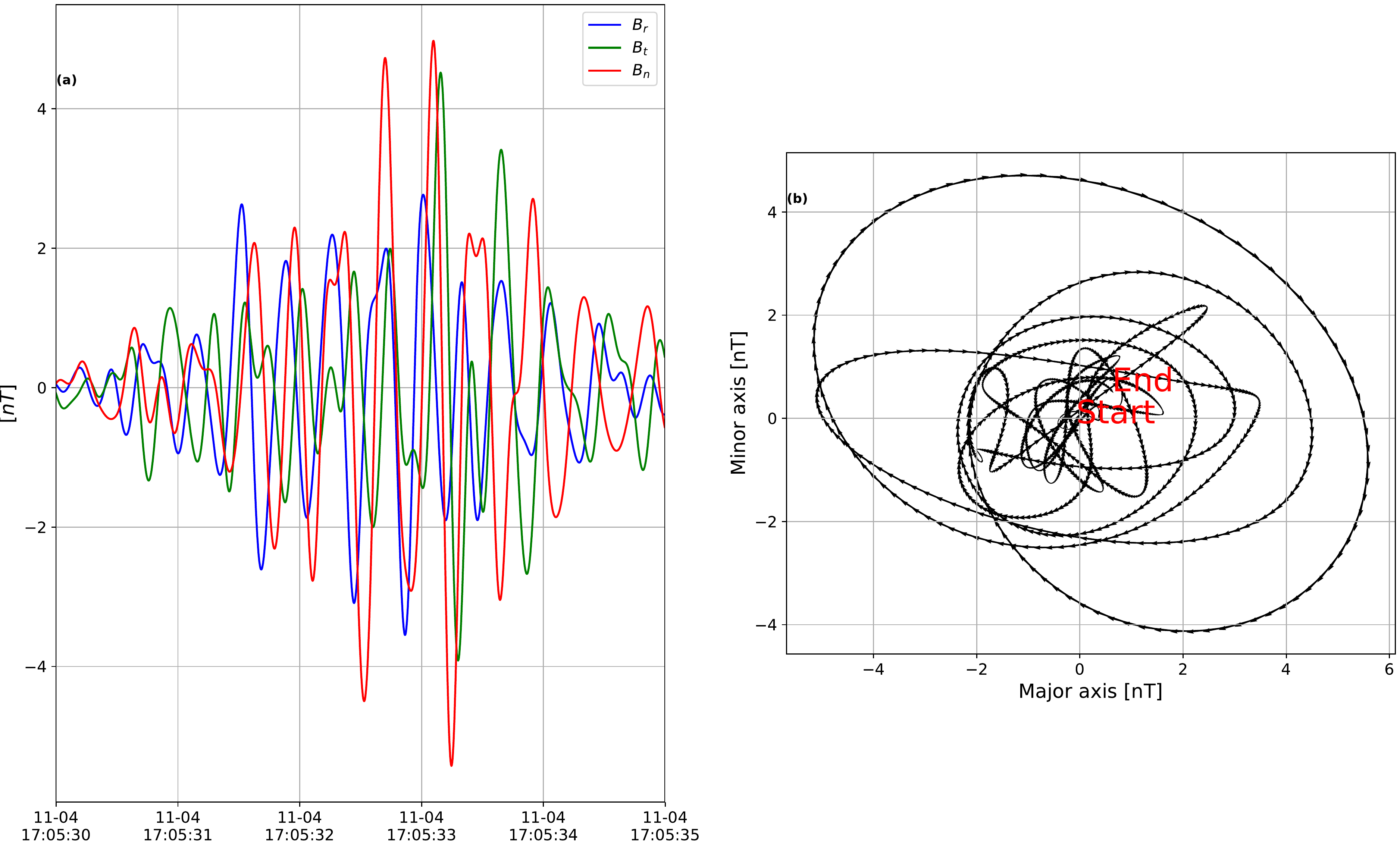} 
    \caption{a) Waveforms of magnetic field components in the RTN frame recorded in a close vicinity of the leading edge of the structure shown in Figure~\ref{fig:fig_9}. b) Hodograph of the magnetic field in the plane transverse to the wave normal vector.}
    \label{fig:fig_11}
\end{figure*}

The variations of magnetic field and velocity around leading edge of the
structure are presented in Figure~\ref{fig:fig_10}. Panel~(a) shows the variations
of the magnetic field as registered by the MAG instrument in the RTN
frame. Panel~(b) represents the variations of the magnetic field components
across the boundary corresponding to the largest (blue), intermediate
(green) and the smallest (red) eigenvalues. In order to make this analysis
the MAG data was filtered to remove the fields in the frequency range higher
than 0.2 Hz. The ratio of the second to third eigenvalues is about 2, that
indicates rather large error bars caused by the wave activity around the
boundary. The normal unit vector to the boundary $N_{lead}=[-0.33;0.07;0.94]$
is close to the normal northward direction with the small radial component.
The normal to the boundary component of the magnetic field is rather large $
B_{n}=$ \SI[separate-uncertainty = true]{24.4 \pm 5.9}{nT}. Panel~(c) shows that variations of the velocity
vector are synchronous with the magnetic field. The panel~(d) shows
variations of the component of the velocity along the normal to the boundary
direction, it varies in the range 85-105  \si{km/s}. It was found to be slightly different to magnetic field defined, the
difference is \ang{19}. Figure~\ref{fig:fig_11} shows the waveform of the magnetic
field components registered around the leading edge of the structure by SCM
filtered in the frequency range of 2-5 Hz. The wave $k$-vector was found in \ang{60} with respect to the normal to
the boundary. This indicates that the wave can be considered as a surface
wave \citep{Hollweg1982}. The wave frequency is rather close to the local ion
cyclotron frequency.

\begin{figure*}
	\centering
    \includegraphics[width=\linewidth]{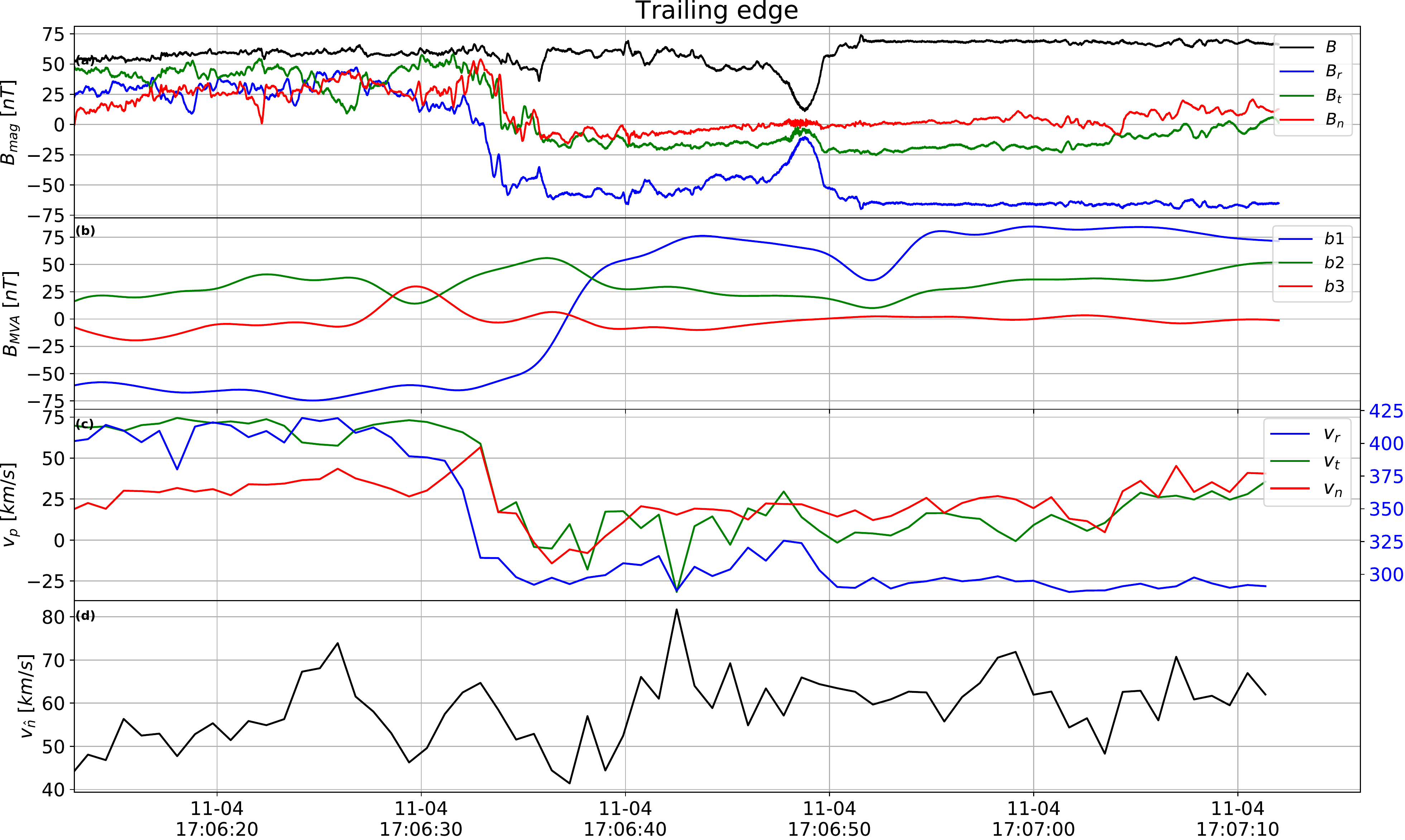} 
    \caption{The magnetic field and plasma bulk flow velocity on the trailing edge of the structure shown in Figure~\ref{fig:fig_9}.}
    \label{fig:fig_12}
\end{figure*}

\begin{figure*}
	\centering
    \includegraphics[width=\linewidth]{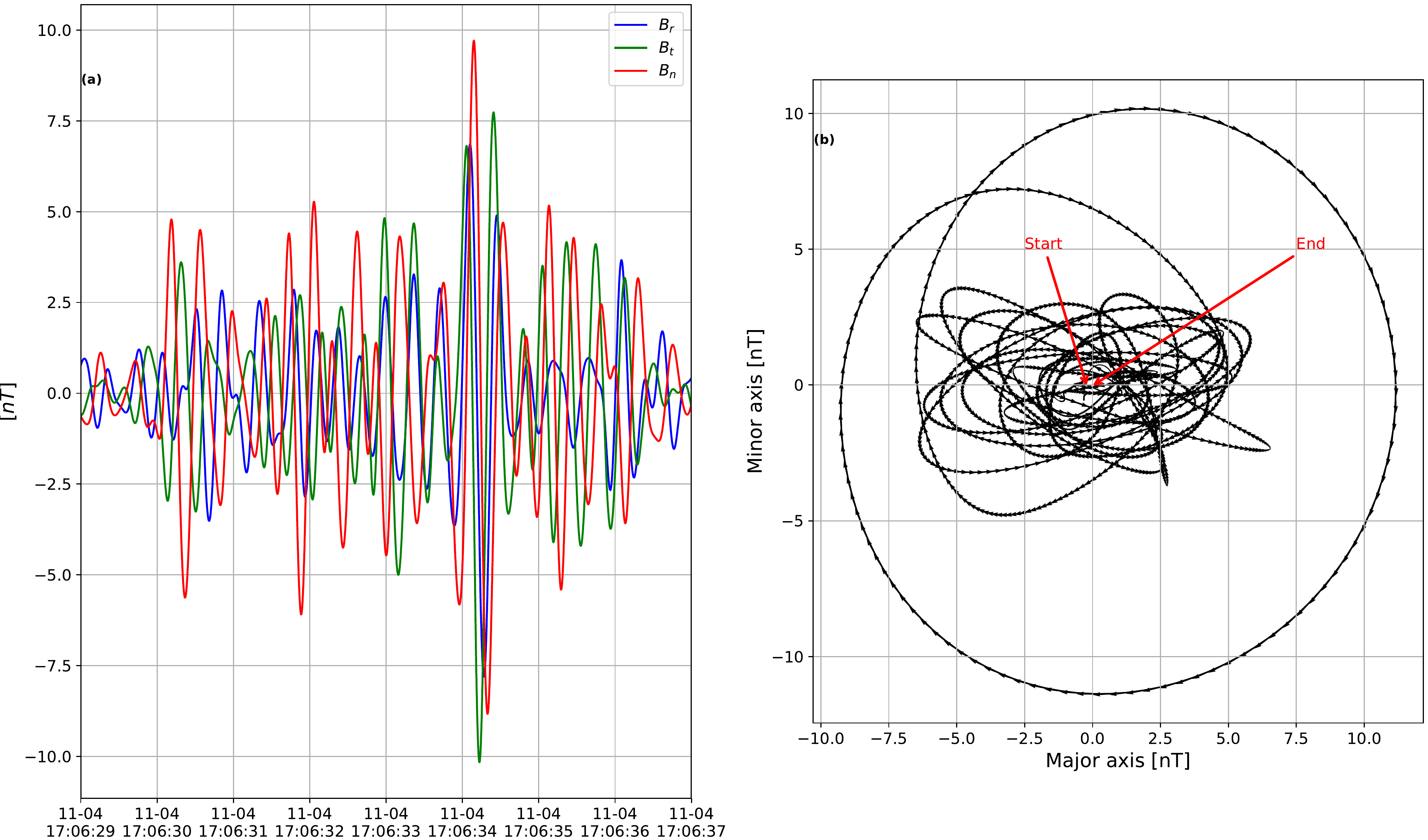} 
    \caption{a) Waveforms of magnetic field components in the RTN frame recorded in a close vicinity of the trailing edge of the structure shown in Figure~\ref{fig:fig_9}. b) Hodograph of the magnetic field in the plane transverse to the wave normal vector.}
    \label{fig:fig_13}
\end{figure*}

Figure~\ref{fig:fig_12} shows variations of the magnetic field and velocity vectors around
the trailing edge boundary. The magnetic
field (panel(a)) returns to the parameters rather close to those before the encounter, the
radial component becomes dominant and normal and tangential become
relatively small. Panel~(b) shows variations of the magnetic field
in the MVA reference frame with the same colors as for the leading
edge. Variations of the velocity vectors (radial is shifted down to 300
 \si{km/s}) are presented on the panel~(c) and on panel~(d) variations of the
normal to the boundary component of the velocity that varies in the range 
40-60 \si{km/s}. However, one should note that there is the important difference
of magnetic field directions before and after encounter: the significant tangential
component of the field while the normal component was sufficiently smaller,
and after the tangential component became practically negligible whereas the
normal component became rather significant. The angle between magnetic field
vectors before and after encounter is approximately \ang{27}, the
angle between velocity vectors is much smaller - about \ang{5.7}.
One can conclude that the structure carries some integral current that
results in important shear of the magnetic field. On Figure~\ref{fig:fig_9} it was
evidenced the presence of quite intense wave activity around both leading
and trailing edges of the structure. The waveform and its hodograph of the wave 
around the trailing edge are presented in Figure~\ref{fig:fig_13}. It is worth 
noting that the angle between $k$-vector of the wave and the normal to the surface 
makes \ang{60.5}. It supports an assumption that this wave is the surface wave. 
It is worth mentioning that there may exist different type of the wave activity around 
the boundary of compressional structure, namely whistler waves, as it was reported 
by \citet{Agapitov2020}.
We present wave characteristics for the trailing edge in more detail in 
Figure~\ref{fig:fig_15}. Panel~(a) shows the waveform of the three 
components of the magnetic field fluctuations obtained
in the frequency range 2-5 Hz. On panel~(b) we show the hodograph of two
largest components of the wave magnetic field that indicate
the wave's close to the circular polarization. The angle between the wave $k$-vector and the normal to
the boundary that is found to be about \ang{80}, so, the wave can
be considered as a surface wave.

\begin{figure}
	\centering
    \includegraphics[width=\linewidth]{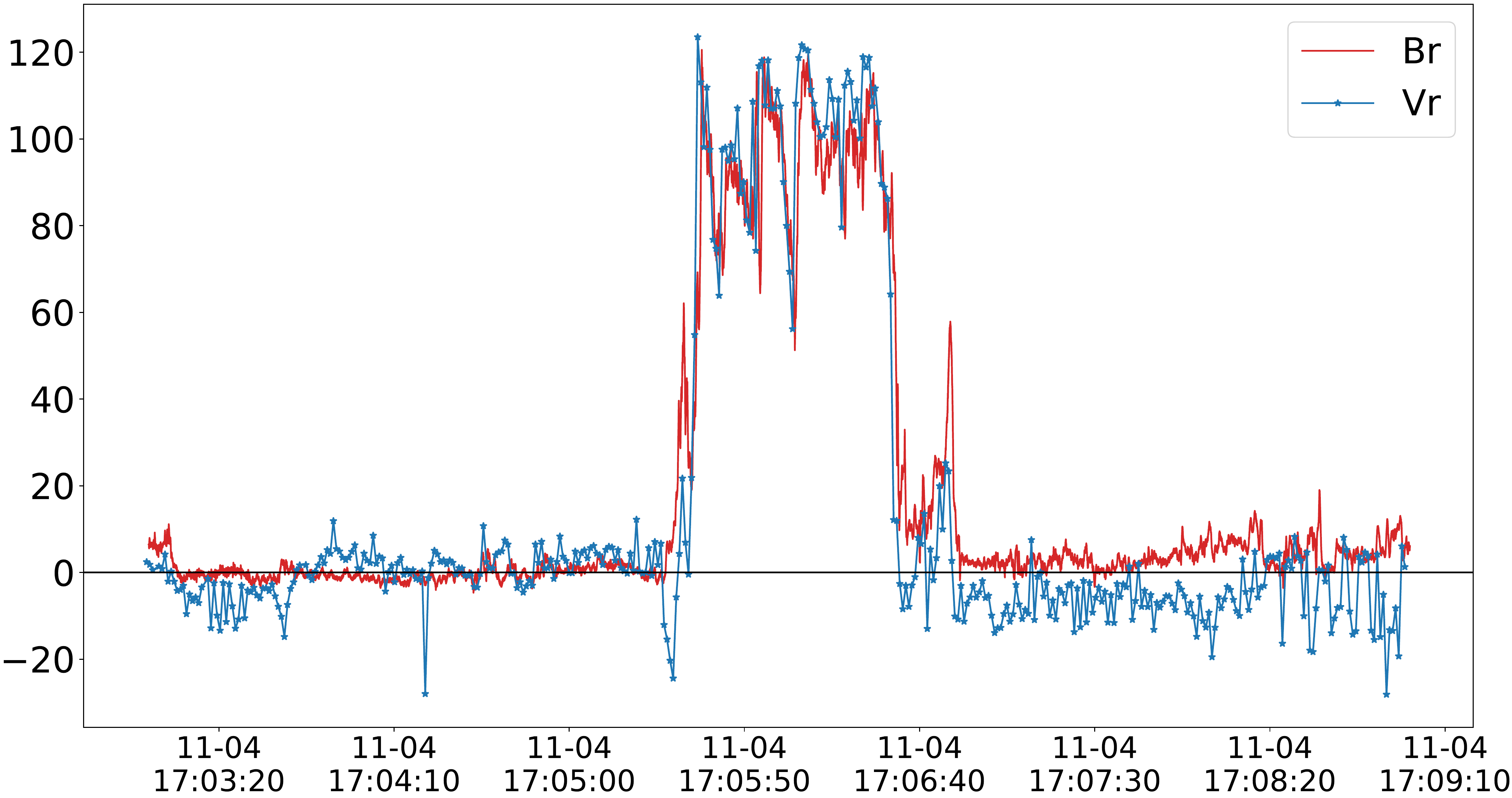} 
    \caption{The radial components of the magnetic field (red) and plasma bulk velocity (blue) during crossing of the structure (Figure~\ref{fig:fig_9}), illustrating close to the linear correlation, i.e. high level of Alfv\'enicity. The Alfv\'enic type of perturbations is on the leading edge, however, the change of components around trailing edge is presumably due to the change of plasma parameters across the structure. The y-scale is the same because the respective mean computed over the before encounter data (17:03 to 17:05:26) has been subtracted.}
    \label{fig:fig_14}
\end{figure}

One of the characteristics used to classify the discontinuities in the solar
wind is Alfv\'enicity. It is manifested in absence of variations of the
magnetic field magnitude and in synchronous variations of the velocity and
magnetic field perturbations related linearly as described in previous
paragraph. We have already mentioned that the magnitude of the magnetic
field manifests quite important variations. This indicates that the
structure may not be considered as pure Alfv\'enic. Correlation between the
decrease of the magnetic field magnitude and ion density provides strong
argument in favor of attribution of this structure to fast magnetosonic
mode. Results of the check of another condition, linear correlation between
perturbations of the magnetic field and velocity vectors are presented in Figure~\ref{fig:fig_14}. It shows variations of the radial component of the magnetic field and of the radial velocity, that undergo the largest changes. On the leading edge of the boundary one can conclude that the correlation is quite
convincing and it is similar to that observed for Alfv\'enic structure.
However, there is a quite important difference of the coefficients of linear
relation between the leading and trailing edges that might arise from the
significant change of the plasma parameters between two boundaries. To
outline the similarity and difference the average values evaluated before
the structure were subtracted. The linear correlation is good for the
leading edge transition, except the region where the local density strongly
increase, but the fitting of the trailing edge boundary shows significant
difference. The normal unit vector to the trailing edge boundary $\mathbf{n}_{tr}=[0.18;-0.59;0.78]$. The component of the
magnetic field orthogonal to the boundary is very close to zero 
$B_n = $\SI[separate-uncertainty = true]{0.0 \pm 1.2}{nT}.
The wave activity of the surface waves around the leading and trailing edges may be characterized by means of evaluation of the of the Poynting flux in the reference frame of plasma inside the tube. The data of measurements of the fields and velocities allow one to to evaluate the z-component of the Poynting flux (it is close to the radial direction). In order to do it, as in previous case the x and y components of the electric and magnetic fields are used in satellite
reference frame. The VxB electric field is evaluated and removed from the data and the Poynting flux is estimated. The results are presented in Figure~\ref{fig:fig_15}. Panel~(a) shows the magnetic field, panel~(b) shows the SCM waveforms, panel~(c) shows the
Poynting flux. The Poynting flux enhancement is observed around both the
leading and trailing edges of the structure, and the Poynting flux is
negative, that corresponds to the direction outward from the Sun.

\begin{figure*}
	\centering
    \includegraphics[width=\linewidth]{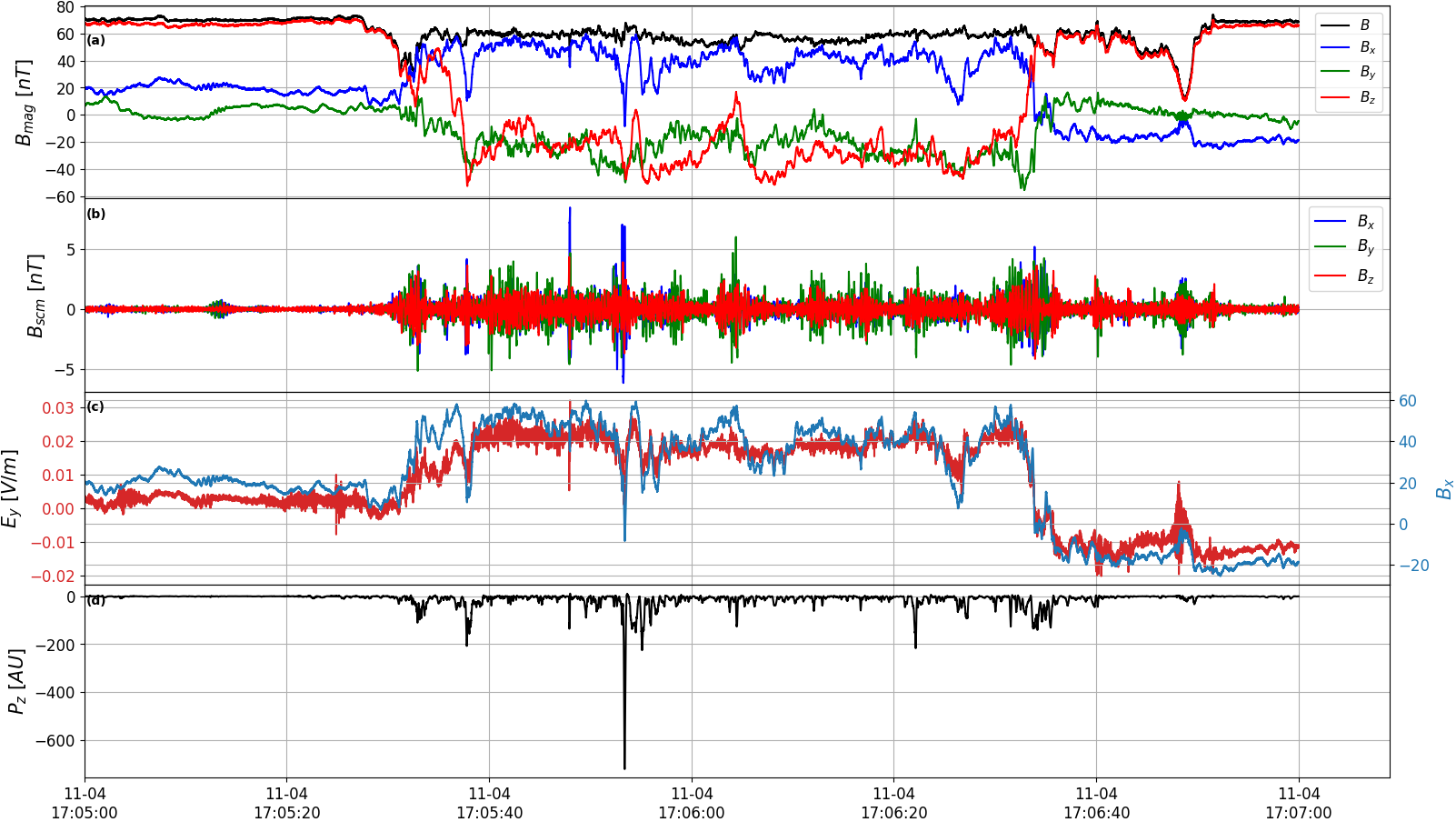} 
    \caption{Poynting flux evaluation for compressional structure. a) Magnetic field measurements by MAG instrument during the structure crossing in satellite xyz reference frame, similarly to Figure 7 the data are presented in the satellite reference frame. The components are colored x in blue, y in green and z in red. b) Magnetic field fluctuations registered by SCM instrument, colors are similar to panel a. c) y-component of the electric field registered by FIELDS suite, in red, and x-component of the magnetic field by MAG instrument in blue, the correlation between variations of components is related to the important electric field induced due $\mathbf{V}\times\mathbf{B}$ plasma motion. These fields components are used together with the x-component of the electric field and y-component of the magnetic field for evaluation of the z-component of the Poynting flux. d) z-component of the Poynting flux during time interval including compressional structure crossing. Quite large fluxes directed along the z-axes are registered. It is worth reminding that z axes is quite close to radial direction and the negative flux is directed outward from the Sun.}
    \label{fig:fig_15}
\end{figure*}

The characteristic spatial variations of the system around and inside the
structure similarly to previous case are much larger than characteristic
scales in plasma, the Larmor radius and ion inertial scale, thus the analysis
of plasma flow motions admits similar approach.  One can separate the total
motion to the flow motion along the magnetic field lines and orthogonal
magnetic field lines motion of the magnetic field tubes. Applying similar
procedures one can find the velocity parallel to magnetic field before
encounter to be equal to 282.8 \si{km/s}, and magnetic field tubes relative
velocity perpendicular to their direction $\mathbf{V}_{\perp/bef}=[127.7;133.3;21.4]$, $\mid V_{\perp/bef}\mid =185.8$ \si{km/s}, here index ${\perp/bef}$ notifies the perpendicular to the magnetic field direction  component of the velocity of plasma before the structure.
Supposing that the satellite crosses a tube-like structure and let us assume
that it is much longer along its axes than in perpendicular directions. For
the sake of simplicity it may be considered as having cylindrical cross
section. Under such assumptions one can find that the flow velocity along the
magnetic field is equal to $\mid V_{par/in}\mid =260.4~\si{km/s} $, and the
orthogonal velocity of the tube may be estimated to be equal to $\mathbf{V}%
_{\perp/in}=[274.4;-132.2;-77.2]$, and $\mid V_{\perp in}\mid$ =314.2~\si{km/s}.
The angle between average magnetic fields inside the structure and outside
before the encounter is \ang{72.2}. Taking estimates of the normal to
the boundary component of the velocity at leading and trailing edges of the
structure one can evaluate the length of the chord corresponding to path of
the satellite as $L\simeq \frac{1}{2}(V_{n/lead}$ $+V_{n/out})\delta t=\SI{7000}{km}
$. The evaluation of the angle between the tube velocity and normal to the
structure shows that the satellite enters inside the cylinder where the
velocity vector makes an angle with the normal to the cylinder of
approximately \ang{56} and quits it with the angle \ang{78}.
These estimates are rather rough but they provide an idea that the crossing
chord is of the order of diameter of the tube, thus the estimate of the
scale above gives reasonable size of it. It is worth noting as in previous
case that the structure supposedly carries some integral current resulting
to significant shear of the magnetic field, the fields before and after it
make an angle of \ang{27}.


\subsection{Event 3 - Switchback with full reversal of the magnetic field}
\label{sec:event3}

The satellite crossed the structure at 5 of November from 04:27:40 UT to
04:42:05. The basic parameters of the plasma during this encounter are
presented in Table~\ref{table:table_3}. 

\begin{deluxetable*}{cccc}
\tablecaption{Major plasma parameters for the radial component reversal structure
\label{table:table_3}}
\tablewidth{0pt}
\tablehead{
\colhead{Parameter} & \colhead{Before encounter} & \colhead{Inside structure} & \colhead{After encounter} 
}
\startdata
magnetic field vector \si{nT} 
& $\mathbf{B}_{bef}=[-69.4;-9.7;-17.7]$ 
& $\mathbf{B}_{in}=[53.6;51.2;19.6]$
& $\mathbf{B}_{after}=[-72.4;13.9;20.2]$  \\ 
magnetic field magnitude \si{nT} 
& $| \mathbf{B}_{bef} |   =$ \SI{72.2}{nT} 
& $| \mathbf{B}_{in}  |   =$ \SI{76.7}{nT} 
& $| \mathbf{B}_{after} | =$ \SI{76.4}{nT} \\ 
\enddata 
\end{deluxetable*}

The plasma moments by the SWEAP instrument are rather poor for this time
interval. The only plasma parameter available for this event is an electron
density estimated from the Quasi-Thermal Noise (QTN) technique. It varies
between 460 to 530 \si{cm^{-3}} inside the structure.

\begin{figure*}
	\centering
    \includegraphics[width=\linewidth]{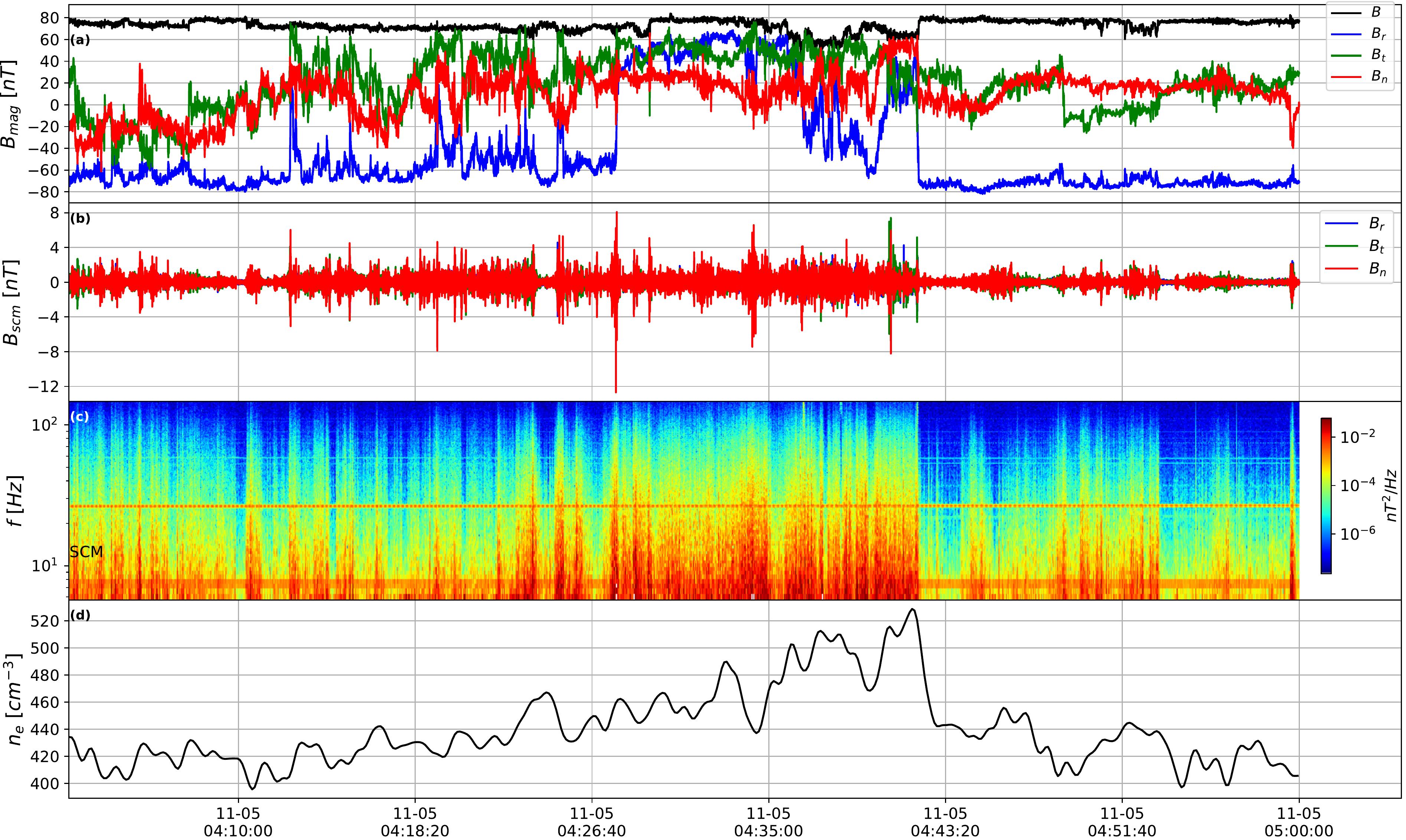} 
    \caption{The structure containing the full reversal of the radial component of the magnetic field in the same format as in Figure~\ref{fig:fig_2}}
    \label{fig:fig_16}
\end{figure*}

Figure~\ref{fig:fig_16} represents variations of the magnetic field, spectra of magnetic fluctuations, and an electron density profile evaluated making use of the QTN 
technique during crossing of the magnetic structure with reversal of the
radial component of the magnetic field. The switchback here consists in
reversal of the radial component of the magnetic field and its return. It
begins at $04:27:40$ when the sharp change of the radial component of the
magnetic field occurs. It results in drastic change of the field magnitude
from \SI{-58}{Hz} to \SI{26}{Hz}. This transition happens in about 10 seconds time
interval. The radial component of the field remains large and positive till $%
04:36:35$, when it begins to return to negative values and reaches a value
of about \SI{-60}{Hz} for short time interval. After that it increases again to
positive but smaller values from $0$ to \SI{45}{Hz} and during time interval $%
04:42:00$ to $04:42:05$ it returns to large negative values, even larger than
before the encounter, of about $70-75$~\si{Hz}. The second panel~b represents the
waveform of the magnetic field fluctuations measured with the SCM instrument and
panel~c represents the power spectral density of the magnetic field
fluctuations evaluated using the SCM instrument. There are many spikes of high
intensity in the frequency range from \SI{5.7}{Hz} to several tens of \si{Hz}. The
panel~d shows an electron density profile obtained from electric field
measurements making use of the QTN technique \citep{moncuquet20}. The profile
presented is smoothed and can not correctly represent the sharp boundaries. Large
scale density variations are of the order of \SI{40}{cm^{-3}}, that is of the
order of $10\%$.

\begin{figure*}
	\centering
    \includegraphics[width=\linewidth]{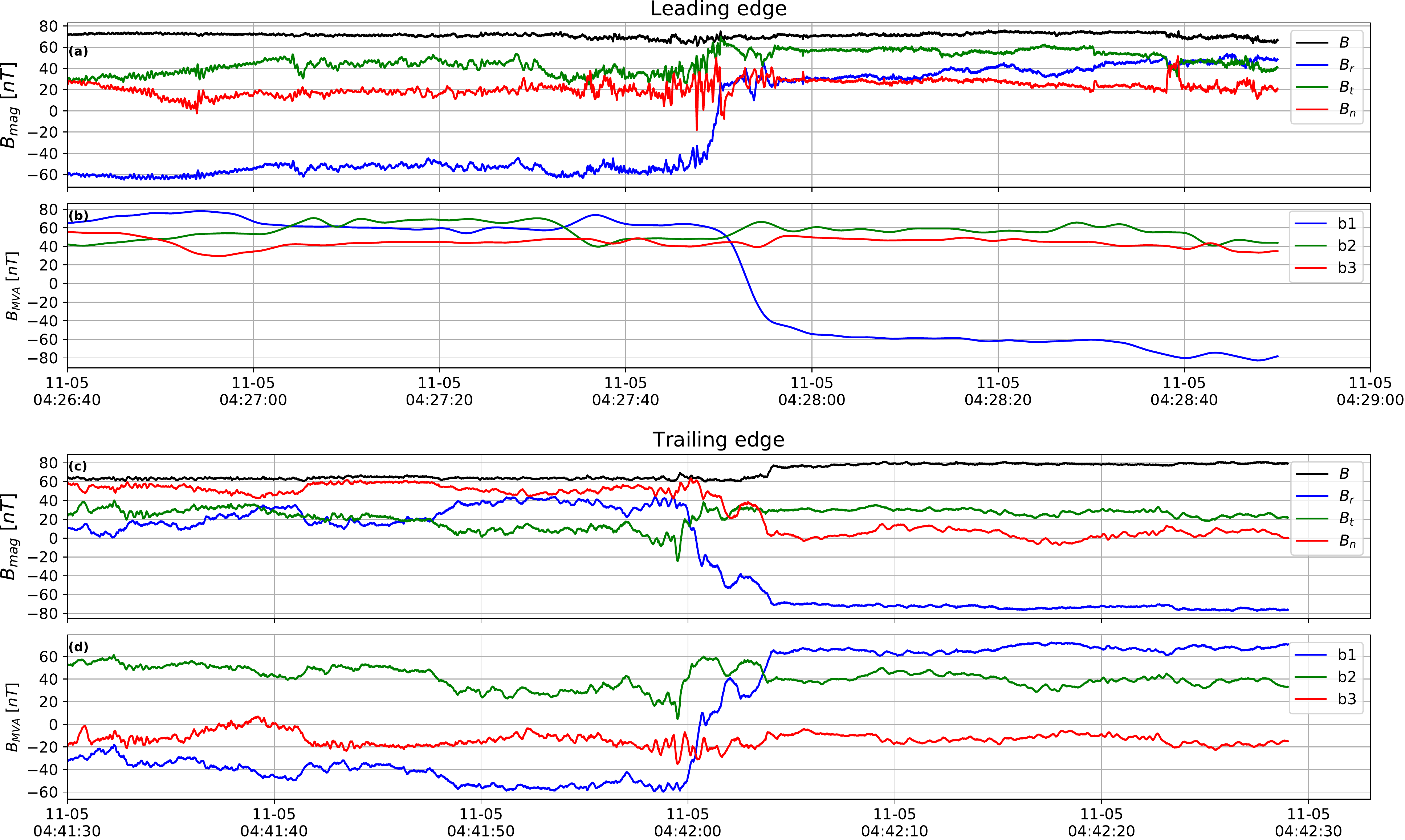}
     \caption{The magnetic field on the structure boundaries: the magnetic field components on the leading edge in the RTN frame (a) and in the MVA frame (b); the magnetic field components on the trailing edge in the RTN frame (c) and in the MVA frame (d).}
    \label{fig:fig_17}
\end{figure*}

In order to characterize the boundaries we carried out the MVA analysis again.
The normal vector to the leading edge is $\mathbf{N=}[-0.12;0.21;0.97]$, it
is very close to the normal direction perpendicular to the equatorial plane.
It is worth noting that in addition to major variation of the radial
component there is small but significant variation of the tangential
component from about 35 to 70~nT. In order to illustrate the field
dynamics across the boundary we present in Figure~\ref{fig:fig_17} a more detailed view of the time variation of the magnetic field around the leading and trailing
edges. The low pass filtered ($<$\SI{0.2}{Hz}) components of the magnetic field are presented on the panel~a of Figure~\ref{fig:fig_17}.  The magnetic field has significant component along the normal to
the boundary of the leading edge 
$B_N =$\SI[separate-uncertainty = true]{-43.9 \pm 5.3}{nT}. Time
variations of the magnetic field components in MVA reference frame are shown
on the panel~b of the same Figure. On the panels presenting the results of the MVA analysis we
use the same colours as for previous events. 
Similar analysis applied to the trailing edge boundary shows that the normal
vector to it $\mathbf{N=}(0.38,0.65,-0.66)$ is mainly in the
normal-tangential plane with small radial component and makes an elevation
angle in tn plane close to \ang{45}. The normal component of
the magnetic field on this boundary is sufficiently smaller than at the
leading edge, it is equal to $B_N =$\SI[separate-uncertainty = true]{-13.3 \pm 5.8}{nT}. It is
worth noting that several of structures we analyzed have large normal
component of the magnetic field on the one boundary and relatively small on
the other. The magnetic field variations for the trailing edge are shown on
the panel~c of the same Figure~\ref{fig:fig_17}, and the variations of the eigenvectors corresponding to largest, intermediate and smallest eigenvalues of the field are shown on the panel~d.
One of the important characteristic features used for determination of the
property of so called "Alfv\'enicity" over the discontinuities consists in the
absence of the variation of the magnetic field magnitude while the
components vary across the boundary. Considering the whole structure one can
observe quite important variations of the magnitude of the magnetic field,
however at the leading edge transition there is no any significant variation
of the magnitude of the magnetic field. The trailing edge transition may be
"non-Alfv\'enic", the magnitude of the field undergoes quite important jump
from about 65~nT to approximately 80~nT. One important observation to be
pointed out emerges from the evaluation of the angle between the magnetic
field before the structure and after its crossing, it makes \ang{35}.
It leads to the suggestion that the currents flowing on the surfaces of the
leading and trailing edges result in the significant shear of the
surrounding magnetic field.
Another important feature of the structures observed is quite high intensity
of the wave activity in the frequency range from 5.7Hz (low limit for the
spectra we adopted) to several tens of Hz, manifested in the spectra of
SCM typically more than order of magnitude higher than in the surrounding
quiet plasma flow, as it was also pointed out by \citet{ddw20}. Unfortunately our analysis for this particular event
is limited by FIELDS instruments measurements only, SWEAP data for it
contain quite poor data set and are not usable for the detailed analysis.


\section{Interpretation of observations and discussion}
\label{sec:interpretation}

Our data analysis may be summarized as follows. The structures we observe are large in comparison with the typical characteristic scales in plasma, such as the Debye length, the ion and electron Larmor radii and the ion and electron inertial lengths, except for their boundaries.
One of the boundaries is of the order of several ion inertial lengths or ion Larmor radius (they are of the same order of magnitude). Time variations of macroscopic motions of plasma are also slow with respect to the ion gyroperiod, that is the largest characteristic time scale of particle motions. The only exception is the wave activity in the vicinity of the boundaries that we shall discuss later.
We examined the properties of three selected structures that we consider as representatives of three different groups, a) Alfv\'enic type structures where magnetic field variations occur without change of the magnitude of the magnetic field, b) compressional type structure where the magnetic field magnitude varies and is accompanied/associated with the variations of the plasma density, c) full reversal of the radial component of the magnetic field, but similar to Alfv\'enic type structure. 
To determine the "local design" of the structures we carried out an analysis of the plasma motions and fields inside and around the structures with special attention to their boundaries. Since the plasma is fully ionized, collisionless and characteristic scales of spatial variations are large, one can treat its macroscopic motions as "frozen in", except its boundaries. This suggests the separation of its motion into two parts, the flow motion of the plasma that is directed along the magnetic field, and the motion of the magnetic field tubes in the direction perpendicular to their axes. Plasma characteristics inside and outside the structures are found to be rather different, and there are several parameters making this difference. Those are: the direction of the
magnetic field, the plasma density, the flow velocity and ion beta, the ratio of the ion thermal pressure to the magnetic pressure, that was found to be larger than one inside the structures and significantly smaller outside.
The magnetic field before the structure and after it manifests significant change. The angles between the fields before and after are larger than \ang{25}  (we show it for our selected examples, but we checked it for about 20 other events), that results in magnetic fields difference of the order of 30 \si{nT}, thus the structures themselves may be considered as the system of current sheets that carries some total current. This change is sufficiently smaller than the difference of fields inside and outside the structure, but it may not be neglected. 
In order to characterize the parameters and geometry of the structures we performed MVA analysis of the magnetic field variations of the boundary current sheets and determined their characteristics (it was done for about 20 other structures but the statistical study will be presented in a separate publication). This analysis allows one to estimate the magnetic field jump and the characteristic direction and the scale of the magnetic field change through the boundaries. 
Now we shall use these characteristics to evaluate the distribution of currents across the boundaries. The knowledge of the direction of the magnetic field inside the tube that coincides with its axes allows also to separate the current observed on both the leading and trailing edges of the structure onto tube-aligned and azimuthal. The azimuthal current provides a diamagnetic effect that compensates magnetic field difference along the tube axes inside and outside. Tube aligned current in its turn flow on both sides of the tube surface in opposite directions and may
lead to the shear of the magnetic field outside the structure. Our analysis reveals the following characteristics for three structures we presented.


\subsection{Alfv\'enic structure}

\begin{figure}
	\centering
    \includegraphics[width=\linewidth,trim={3cm 6cm 3cm 6cm},clip]{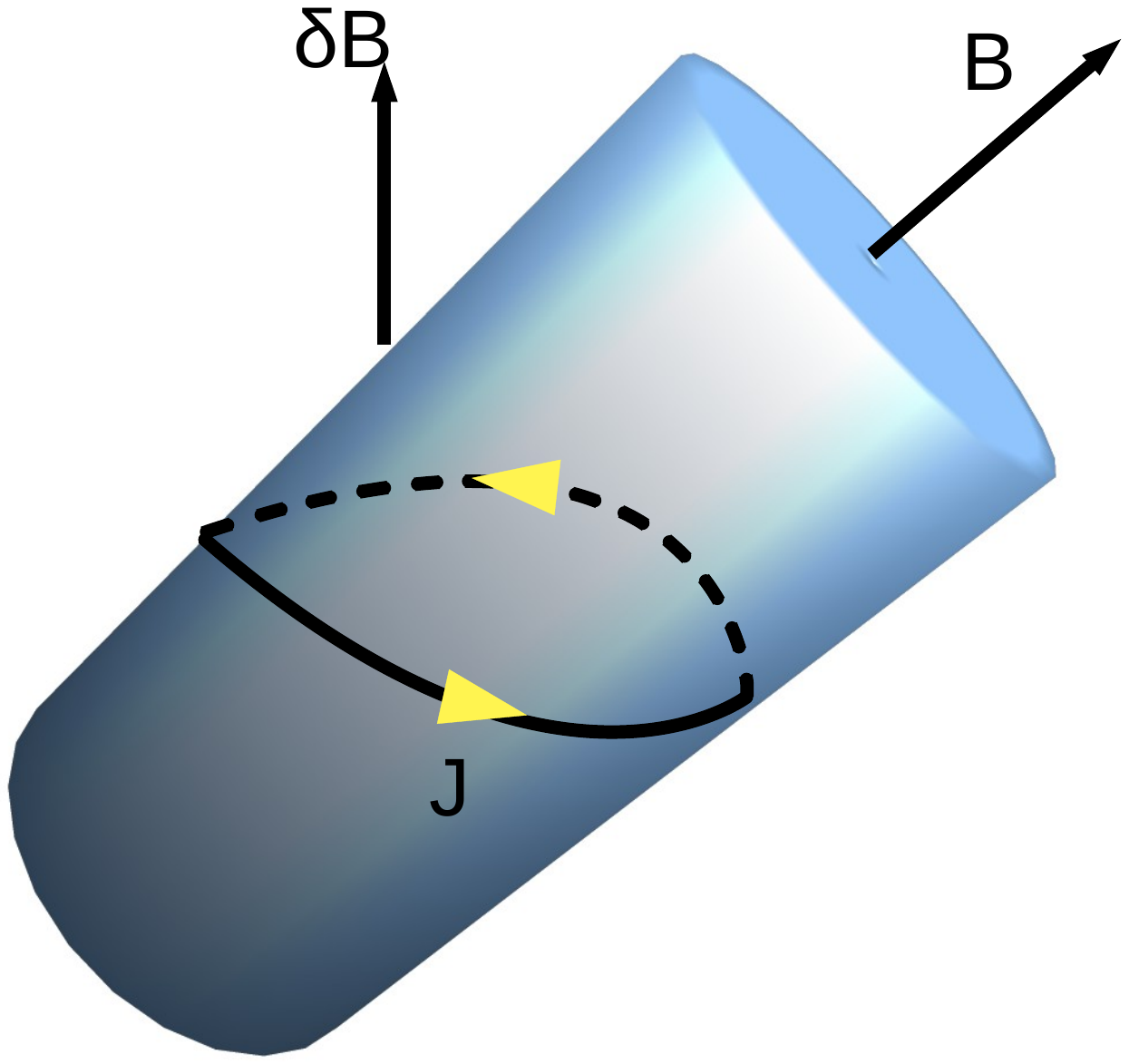} 
    \caption{The geometry of the current system on the surface of the magnetic tube. Here $\mathbf{B}$ is the magnetic field vector directed along the axes of the magnetic tube;  $\delta\mathbf{B} = \mathbf{B}_{inside} – \mathbf{B}_{outside}$ is the difference of magnetic field vectors inside and outside the tube. The elementary current loop on the surface in the plane perpendicular to $\delta \mathbf{B}$.}
    \label{fig:fig_18}
\end{figure}

The leading edge of the structure, more precisely the major jump of the field (it may be called ramp similarly to shocks) was crossed on 6 November from 23:32:48.4 to 23:32:49.6, i.e. duration of crossing is approximately 1.2 sec (see details on Figure~\ref{fig:fig_5} panels a and b). The jump of the magnetic field during ramp crossing was about 75 \si{nT}, but it was additional smaller jump at 23:32:56.8 of the duration of 0.6 seconds and only then the final value of the field inside the tube was achieved. For the sake of simplicity of estimate we shall take it as one single jump from outside to inside value of the field of the duration of 1.8 seconds. The average normal velocity during this period was about 25 \si{km/s} (panel d of Figure~\ref{fig:fig_5}), thus the characteristic spatial scale of the layer is about $\delta L\simeq 45km$. In order to evaluate an average magnetic field before encounter we choose an interval from 23:29:10 to 23:30:49 when the variations of the fields components are relatively small (less than $10 \%$) and filter the data keeping only frequencies less than 1 Hz. Then we calculate an average magnetic field on this interval before encounter outside the structure. To evaluate the average magnetic field inside the structure we take an interval from 23:33:51 to 23:38:12 that is slightly distant from both boundaries and proceed in similar way. We proceed in the same way to compute the average field after the encounter in the interval 23:40:20-23:42:56. Average magnetic field vector outside before encounter is $<\mathbf{B}_{before}> = [-75.8;-17.5;22.5]~\si{nT}$, and inside $<\mathbf {B}_{inside}> = [-46.1;27.5;-63.6]~\si{nT}$. This results in total jump of magnetic field vector $<\mathbf{\delta B>}~=[29. 7;45.0;-86.1]$, with magnitude $\mid \mathbf{\delta B\mid }=\SI{101.6}{nT}$. 
It is worth noting that the angle between the normal to the boundary and upstream magnetic field before encounter makes \ang{82}. The normal component of the magnetic field is found to be $B_n =$\SI[separate-uncertainty = true]{-2.5 \pm 3.2}{nT} that makes it reasonable to speak about quasiperpendicular boundary layer, or in terms of discontinuities TD. The evaluation of the current density results in $j\simeq \frac{\delta B}{\mu _{0}\delta L}\simeq 1800%
\frac{nA}{m^{2}}$. The current flows along the narrow surface of the tube. Its thickness is of the order of 4-5 ion Larmor radii. The current makes an angle of \ang{23} with the direction of the magnetic field outside the tube, and an angle of \ang{30} with the axes of the magnetic tube. Thus the current density along the tube axes, that we might call the tube aligned current density is approximately $j_{par}\simeq 1530\frac{nA}{m^{2}}$, and the azimuthal component $j_{az}\simeq 950\frac{nA}{m^{2}}$. The difference of  magnetic field between the two sides of the structure is significantly smaller than the difference between the field inside and outside the structure. It justifies a suggestion that in first order approximation the current closure occurs on the same surface of the tube. Surprisingly the tube aligned component of the current is significantly larger than azimuthal which signifies that the current circuit element should be placed on very oblique ellipsoid. It is easy to understand because this current should provide the magnetic field equal to the difference of magnetic field inside and outside the structure. If we consider that this structure could be produced by the current loop it should flow in the plane perpendicular to the vector equal to the difference of the two magnetic field vectors.
The trailing edge crossing time is also about two seconds from 23:39:02 to 23:39:04, but the normal component of the velocity is about 220 \si{km/s} as one can see on Figure~\ref{fig:fig_6}, almost ten times higher than for leading edge, thus the thickness of the trailing edge current sheet is about ten times larger, about 440 km. Its width is sufficiently larger than any characteristic scale of the plasma. The boundary is significantly broader and more diffuse, but the thickness of the layer remains much smaller than the size of the structure. The difference of the average magnetic field after encounter and inside the structure is:

\begin{equation*} 
\delta \mathbf{B}=[28.\,\allowbreak 3;8.7;86.6]~\si{nT}\text{ , }\mid \delta B\mid =91.5~\si{nT}
\end{equation*}%
The current density may be evaluated to be 
\begin{equation*}
\frac{\delta B}{\mu _{0}\delta L}\simeq 165\frac{nA}{m^{2}}
\end{equation*}%
It is an order of magnitude smaller than the one at the leading edge. The current has a direction almost opposite to the current on the surface of the leading edge, the angle between the direction of the tube axes and the direction of the current makes \ang{33}, and tube aligned and azimuthal components of the current density have magnitudes as follows 
\begin{equation*}
j_{par}=139\frac{nA}{m^{2}};j_{az}=89\frac{nA}{m^{2}}.
\end{equation*}
The cut of the tube corresponding to the surface where the current flows will present an ellipsoid very oblique with respect to the tube axes, the angle the largest axes of ellipsoid makes with the tube axes is about \ang{33}. The sketch drawing showing such current element is presented on Figure~\ref{fig:fig_18}.

It is worth noting that the angle between the direction of the current flow and an average magnetic field after encounter makes \ang{34}. 
An important feature of the system to be mentioned that the magnetic fields before and after encounter make an angle of \ang{30} that results in the difference of the fields $\mathbf{\Delta B}=[1.35;36.3;0.5]~\si{nT}$. Its magnitude is almost the same as the tangential component, as the vector is directed almost exactly in tangential direction. It signifies that the structure
itself represents a global current sheet. The precision of our measurements does not allow to correctly evaluate its detailed characteristics However an interesting estimate about the
direction of the global magnetic field may be found from the relative directions of the vectors of  $\mathbf{\Delta B}$ and the axes of the magnetic tube. One can infer that the global current should flow under quite small angle to the axes of the tube \ang{17}. 
The currents at leading and trailing edges need an application of some electric field. As it is known \citep{Braginskii1965} the role of the "effective electric field" may play the gradient of the ion temperature and ion pressure. As it was shown on Figure~\ref{fig:fig_4} both gradients are present and quite large. For the narrow current sheet at the leading edge the current
may be associated with the difference of ion and electron dynamics due to the relatively small scale of the layer in comparison with the ion Larmor radius being just several times larger it. From the other hand for thick layer observed at the trailing edge the current may be determined by the presence of some anomalous collisions due to the wave activity. As it was shown on
Figure 6 there is quite intense wave activity at this boundary due to the presence of the surface wave having an angle with the boundary normal of \ang{60.5}. The wave may affect parallel to magnetic field motions of electrons, as the field aligned currents are carried preferably by electrons. It may also lead to an important energy and impulse exchange between ions having different energies resulting to azimuthal deviations of ion trajectories to provide important ion currents orthogonal to magnetic field lines. 
The wave electromagnetic field has smaller scale than the boundary layer that may make the boundary surface rippled. The wave activity may also significantly enhance diffusion processes through the boundary that will lead to its widening.

\subsection{Compressional structure}

This structure was traversed during significantly shorter time period and
our estimate above showed that its characteristic cross-scale of the tube is
about 7000~km. The leading edge of it was crossed in time interval from 
17:05:26 to 17:05:37 that took 11~seconds with rather high normal velocity
about 90~\si{km/s}, so the thickness of the current sheet layer is quite large,
about 1000~km, thus it is not really a sharp boundary but rather diffuse.
In terms of relative angle between the upstream magnetic field and normal to
boundary the angle makes \ang{71.6}, the transition remains
quasiperpendicular but the relative velocity of the incoming flow is smaller
than the fast magnetosonic speed thus the effect of the "shocking" of the
flow consists only in local increase of the plasma density but not in the
scale of the current sheet. The normal component of the magnetic field is
evaluated to be \SI{24.8}{nT}, thus the discontinuity is in between the TD and/or
RD. The total jump of the magnetic field between the average upstream flow
and average flow inside the tube is expressed as a vector $\delta \mathbf{B} = [95.43;25.71;23.24]~\si{nT}$, its magnitude is \SI{101.5}{nT}. The
current density inside the layer may be evaluated as: 
\begin{equation*}
j\simeq \frac{\delta B}{\mu _{0}L}= \SI{115}{nA/m^2}.
\end{equation*}

The current flow direction makes an angle of \ang{58.4} with the
direction of the tube, thus the projection of the current density to
tube-aligned direction $j_{along} = \SI{62.6}{nA/m^{2}}$, and the projection
to azimuthal direction is about $j_{az} = \SI{98.6}{nA/m^{2}}$. For this
structure the azimuthal current is about $50\%$ higher than the tube
aligned. The angle between current flow direction and upstream magnetic
field makes \ang{60.7}, the field aligned current with respect to the
external magnetic field is also significantly smaller than the azimuthal one. The
effective electric field similarly to previous structure is ensured by the
gradients of the ion pressure and increase of the ion temperature. Efficient
"collisions" similar to above described may be ensured by the wave activity
shown on Figure~\ref{fig:fig_11}. The angle between the $k$-vector of the wave and the
normal to the boundary estimated by MVA is \ang{62.5}.
The analysis of the characteristics of the trailing edge of this structure
(see Figure~\ref{fig:fig_12}) shows that the crossing duration was about 10 seconds to
begin from 17:06:26 till 16:06:36. The average velocity along the normal to
the surface was about $55 \si{km/s}$, thus the characteristic width of the
current sheet layer was about 550~km. The increment vector of the magnetic
field 
\begin{equation*}
\delta\mathbf{B}= \mathbf{B}_{after}-\mathbf{B}_{inside}=[-92.3;-55.3;8.45-13.2] \si{nT}
\end{equation*}

\begin{equation*}
|\delta\mathbf{B}| = \SI{108.4}{nT}.
\end{equation*}

An estimate of the current density gives the following result: 
\begin{equation*}
j = \SI{157}{nA/m^{2}}.
\end{equation*}%
The normal component of the magnetic field is evaluated to be close to zero.
The angle of the current axes with the axes of the tube is \ang{64.2},
according to the following estimates for the parallel and azimuthal
components of the current density: 
\begin{equation*}
j_{par}= \SI{69.1}{nA/m^{2}}, \qquad j_{az}= \SI{141.3}{nA/m^{2}}.
\end{equation*}

Here the azimuthal component of the current is about twice larger than
normal component of the current. The current direction is approximately
opposite to the current on the leading edge of the structure. An assumption that the current closure approximately takes place on the surface is validated similarly to previous case. One can represent again the current element as quite oblique ellipsoid  crossing the structure under corresponding angle of inclination that would generate the magnetic field perpendicular to it. 
The source of the electric field is again associated with the gradient of the ion pressure and ion density between the plasma inside and outside the structure. The effective anomalous collisions that lead to separation of electrons and ions motion are presumably determined by the surface wave
activity as in previous cases. Evaluating the global current direction ensuring the shear of the magnetic 
field one can note that the inclination angle of the current element to the magnetic field axes
makes an angle of \ang{35.5}. 
Our data do not allow us to come to any detailed analysis of currents for the first event, because we have no information about the thickness of the current sheet. However we can evaluate the angle of inclination of the global current with the axes of the magnetic tube, and we found it to be rather large, about \ang{58}. 
The major characteristics of two structures described above are summarized in Table~\ref{table:table_4}. 

\begin{deluxetable*}{ccc}
\tablecaption{Characteristics of boundary layers of the structures}
\label{table:table_4}
\tablewidth{0pt}
\tablehead{
\colhead{Parameter} & \colhead{Alfv\'enic structure} & \colhead{Compressional structure} 
}
\startdata
Characteristic perpendicular size of structure \si{km} & 50000  & 7000 \\ 
Leading edge  (LE) thickness \si{km} & 45 & 1000 \\ 
LE current density \si{nA/m^{2}} & 1800 & 115 \\ 
LE azimuthal current density \si{nA/m^{2}} & 950 & 115 \\ 
LE tube aligned current density \si{ nA/m^{2}} & 1530 & 62 \\ 
LE current density angle with tube axes \si{(degrees)} & 29.8 & 58.4 \\
Trailing edge (TE) thickness \si{km} & 440 & 550 \\ 
TE current density \si{nA/m^{2}} & 165 & 157 \\ 
TE azimuthal current density \si{nA/m^{2}} & 89 & 143 \\ 
TE tube aligned current density \si{ nA/m^{2}} & 139 & 69.1 \\ 
TE current density angle with tube axes \si{(degrees)} & 32.7 & 62.4 \\
\enddata
\end{deluxetable*}

Until now our analysis was dedicated to the local properties of two typical relatively small magnetic structures or switchbacks. The longer duration structures have often sufficiently more complicated form including multiple substructures, that is why we have selected rather pure and simplest cases to shed light to the local "design" of the tubes and to demonstrate the
usefulness of methods that may be applicable for more complicated events also. Our analysis certainly needs to be verified by statistical study that will be presented in separate publication. Our study provides a reasonable support to interpretation of the observed phenomena in terms of moving kinked magnetic tubes. Our approach is local thus it does not allow to address many important issues, in particular:
where the tubes are originated? what causes the twisting of the tubes, is it forced by the collision with the solar wind perturbations or it is produced due to the micro-instability development? do the foot point motions play any role in the formation of the magnetic tubes twists by means of meandering of the field lines as was considered by \citet{Pommois2002}, or may it be produced by some instability like firehose or mirror?  Further statistical studies on different distances closer to the Sun can provide more information to get answers to these questions. 
However, several features of the structures, namely, their similarity with the twisted field lines and enhanced parallel pressure inside indicate that they are very similar to twisted magnetic structures formed as a result of the Firehose instability. Hereafter we consider some of its characteristics that may provide us with some keys leading to interpretation of observed physical phenomena.

\subsection{Firehose instability}

Our analysis of the magnetic structures is based on \textit{in situ} measurements during crossing of them by the satellite. In larger scale it is natural to suggest that the observed magnetic field deviations from the surrounding field lines correspond to a deformation of the magnetic field tube from straight line to twisted or kink type. The overwhelming majority of the flow is directed almost radially from the Sun and the magnetic field lines are also close to the anti-radial direction which supports a strong argument that these magnetic tubes are opened field lines. Enhanced parallel beta inside the structures with respect to surrounding plasma raises the question: may the evolution of the initially straight and almost radial magnetic tubes to twisted and deformed configuration be caused by an instability of anisotropic ion distribution? The well-known instability that may result in such magnetic tube deformation is firehose instability discovered in late 50th by \citet{Rosenbluth1956}, \citet{Vedenov&Sagdeev1958} and \citet{Parker58}. They determined the necessary and sufficient condition for the ion anisotropic instability to occur when the condition $\beta_{\parallel i} - \beta_{\bot i} > 2$ is satisfied.  It was shown that due to it here may grow two types of waves: the shear incompressible Alfv\'en waves and compressible Alfv\'en waves. These instabilities were studied in detail by many authors in different versions of Chew-Goldberg-Low (CGL) MHD and kinetic approximations \citep{ShapiroandShevchenko1964,KennelandSagdeev1967a, KennelandSagdeev1967b, KennelandScarf1968, BerezinandSagdeev1969, Berezin1972, BerezinandVshivkov1976, Gary1993, QuestandShapiro1996, Gary1998, Horton2004}. Statistical studies of the radial evolution of the solar wind from 0.3 to 1.0 AU showed that the ion distribution often consists of a core and a beam with relative velocity of the order of Alfv\'en speed \citep{Marsch_d, Marsch2012}. Simple analysis of the radial dependence of the plasma parameters based on double adiabatic invariants (CGL) approximation assuming that the magnetic field is radial and the density dependence upon radius $\sim R^{-2}$, leads to radial dependence of perpendicular temperature $\sim R^{-2}$ and parallel temperature constant \citep{Matteini2011}. This implies that parallel beta increases with the distance. However, experimental data are clearly in disagreement with the results of such analysis. \citet{Kasper2003} have shown that the temperature anisotropy is indeed constrained by thresholds of the microinstabilities. \citet{Matteini2011} concluded that there should exist parallel cooling and perpendicular heating. A strong support to the presence of these processes is provided by statistical analyses of the characteristic parallel and perpendicular temperatures and parameters of beta in a slow solar wind. It unambiguously shows that the distribution is limited in some area by curves corresponding to thresholds of firehose and mirror instabilities \citep{Bale2009, Hellinger2006}. Kinetic study of these instabilities including Hall and Finite Larmor Radius (FLR) corrections resulted in more accurate estimate of the threshold, it was found to be $\beta_{\parallel i} - \beta_{\bot i} \gtrsim 1.4$ and to discovery of the instability of oblique Alfv\'en wave mode \citep{HellingerandMatsumoto2000, HellingerandMatsumoto2001, WangandHau2003, HauandWang2007}. It was also found that the nonlinear saturation of the shear incompressible Alfv\'en wave (it is also called whistler type) instability occurs in quasi-linear manner, namely the wave amplitude reaches some finite value that corresponds to a decrease of the anisotropy. It is worth noting here that the larger amplitudes would correspond to larger angular deviations and deformations of the field similar to those we observe inside the structures. This value is determined by the transition to the state of marginal stability, the instability is locked. After that the amplitude of wave slightly decreases and remains on this level slowly damping. This level corresponds to the condition of the marginal stability \citep{QuestandShapiro1996,Gary1998,HellingerandMatsumoto2000,HellingerandMatsumoto2001,Matteini2006,Schekochihin2010}. On the other hand the evolution of the compressible wave is slightly different, the wave initially grows to some value and then decreases to some level of saturation, while the ratio of the perpendicular to parallel temperature grows to significantly higher level than in the first case. It is reasonable to suggest that the two types of structures we analyzed here may represent the local manifestations of the tube deformations due to these two types of instability. Our observations supposedly correspond to a saturated stage of the instability development. Since the ion anisotropy may be produced due to two different mechanisms: conservation of the adiabatic invariance and energy of particles while the magnetic field decreases with the distance from the Sun, and due to the presence of the ion beam with the relative velocity larger than the Alfv\'en velocity. An important opened question that needs more detailed analysis of ion distribution functions: is the ion distribution monotonous or it consists of a core and a beam or several beams as it was stated by \citet{Marsch_d} and \citet{Marsch2012} who analyzed the ion distributions registered onboard Helios from 0.3 to 1 AU.  Assuming that on some distance closer to the Sun ion anisotropy is provided by the presence of a distribution with a core and a beam it is reasonable to suggest that the energy source for these structures may be provided by the jets in the low corona of the Sun that are often observed on the boundaries of the equatorial coronal holes \citep{Nistico2009,Raouafi2016}. Such an assumption is supported by the long range correlations between structures noted by \citet{ddw20}.  In the low corona they have velocities of the order of \SIrange{250}{400}{km/s} \citep{Nistico2009,Paraschiv2010} and being combined with the bulk flow propagating outwards may ensure the conditions for instability to be satisfied, since the Alfv\'en speed decreases with the radial distance. In such a picture the macroscopic role of the observed magnetic structures, or switchbacks may consist in providing a mechanism of dissipation of energy supplied by the jets generated in the low corona.

\section{Conclusion}
\label{sec:conclusion}
Our analysis of three typical switchbacks with different characteristics suggests that these structures are magnetic flux tubes moving  perpendicularly to their axis. They are filled with a ``frozen in'' plasma that flows along their local axial magnetic field. However, all of them have boundary layers in which the ``frozen in'' conditions are broken and strong currents flow. As a first order approximation these currents are closed on the boundary surface. An important property is their obliqueness with respect to the axis of the magnetic tube. The cross-section surface of the elementary current loop element is perpendicular to the vector representing difference of magnetic fields inside and outside the structure. The surface boundary layer may vary from several tens of kilometers (which corresponds to several ion Larmor radii) to several hundred kilometers (considerably larger than the ion Larmor radius or ion the inertial length). In such circumstances the surface may carry intense surface waves that have large angles with respect to the normal to the surface. 

Another important feature of these switchbacks is the difference of ion beta inside and outside of the structure. This may in turn ensure the presence of the strong ``effective'' electric fields inside this layer associated with gradients of the plasma density and plasma temperature. Wave activity is confirmed by an analysis of the $z$-component of the Poynting flux (where $z$ is close to the radial direction). All these elements allow to the system to remain quasi-stable and to evolve slowly moving through surrounding plasma.This last may also modify the structure due to inhomogeneous forcing. The full reversals quite probably are produced due to deceleration of the deformed elements of the structure by surrounding plasma.

The deflection of the magnetic field  before and after crossing the switchback further reveals the existence of a total current that is carried by the structure.  These deflections can be as large as \ang{30}, which corresponds to changes in the vector magnetic field of about \SIrange{20}{30}{nT}. From these local characteristics we conclude that these magnetic tubes are most likely twisted at larger scales.  Such deformations of magnetic field lines qualitatively resemble marginally stable structures formed as a result of the development of the firehose instability.

\acknowledgements The FIELDS experiment was developed and is operated under NASA contract NNN06AA01C. VK, TD, AL and CF acknowledge financial support of CNES in the frame of Parker Solar Probe grant. VK and OA were supported by NASA grant 80NSSC20K0697. VK and TD are grateful to P. Fergeau, M. Timofeeva, G. Jannet, P. Martin, M. Maksimovic, K. Amsif and F. Gonzalez for their continuous support during the instrumental and scientific activities of the PSP project. SDB acknowledges the support of the Leverhulme Trust Visiting Professorship programme. Parker Solar Probe was designed, built, and is now operated by the Johns Hopkins Applied Physics Laboratory as part of NASA’s Living with a Star (LWS) program (contract NNN06AA01C). Support from the LWS management and technical team has played a critical role in the success of the Parker Solar Probe mission. The data used in this study are available at the NASA Space Physics Data Facility (SPDF), \url{https://spdf.gsfc.nasa.gov}.

\bibliographystyle{aasjournal}                        
\bibliography{paper}

\end{document}